\documentclass[twocolumn]{aastex62}
\usepackage{graphicx}
\usepackage[utf8]{inputenc}
\usepackage{soul}
\usepackage{amssymb}
\usepackage{booktabs} 

\received{December 11, 2017}
\revised{}
\accepted{}
\submitjournal{The Astrophysical Journal}

\shorttitle{The High cadence Transient Survey (HiTS)}
\shortauthors{J. Martínez et al.}

\begin{document}

\title{THE HIGH CADENCE TRANSIENT SURVEY (HITS)\\ Compilation and characterization of light--curve catalogs}

\correspondingauthor{Jorge Martínez-Palomera}
\email{jorgemarpa@ug.uchile.cl}

\author[0000-0002-7395-4935]{Jorge Martínez-Palomera}
\affiliation{Department of Astronomy, Universidad de Chile, Chile}
\affiliation{Center for Mathematical Modeling, Universidad de Chile, Chile}
\affiliation{Millennium Institute of Astrophysics, Santiago, Chile}

\author{Francisco Förster}
\affiliation{Center for Mathematical Modeling, Universidad de Chile, Chile}
\affiliation{Millennium Institute of Astrophysics, Santiago, Chile}

\author{Pavlos Protopapas}
\affiliation{Institute for Applied Computational Science, Harvard University, Cambridge, MA, USA}

\author{Juan Carlos Maureira}
\affiliation{Center for Mathematical Modeling, Universidad de Chile, Chile}

\author{Paulina Lira}
\affiliation{Department of Astronomy, Universidad de Chile, Chile}

\author{Guillermo Cabrera-Vives}
\affiliation{Department of Computer Science, Universidad de Concepción, Chile}
\affiliation{Millennium Institute of Astrophysics, Santiago, Chile}

\author{Pablo Huijse}
\affiliation{Department of Electrical Engineering, Universidad de Chile, Chile}

\author{Lluis Galbany}
\affiliation{PITT PACC, Department of Physics and Astronomy, University of Pittsburgh, Pittsburgh, PA 15260, USA}

\author{Thomas de Jaeger}
\affiliation{Department of Astronomy, University of California Berkeley, USA}

\author{Santiago González-Gaitán}
\affiliation{CENTRA, Instituto Técnico Superior, Universidade de Lisboa, Portugal}
\affiliation{Center for Mathematical Modeling, Universidad de Chile, Chile}

\author{Gustavo Medina}
\affiliation{Department of Astronomy, Universidad de Chile, Chile}

\author{Giuliano Pignata}
\affiliation{Departamento de Ciencias Físicas, Universidad Andres Bello, Avda. República 252, Santiago, 8320000, Chile}
\affiliation{Millennium Institute of Astrophysics, Santiago, Chile}

\author{Jaime San Martín}
\affiliation{Center for Mathematical Modeling, Universidad de Chile, Chile}

\author{Mario Hamuy}
\affiliation{Department of Astronomy, Universidad de Chile, Chile}
\affiliation{Millennium Institute of Astrophysics, Santiago, Chile}

\author{Ricardo R. Mu\~noz}
\affiliation{Department of Astronomy, Universidad de Chile, Chile}

\begin{abstract}

 The High Cadence Transient Survey (HiTS) aims to discover and study transient objects with characteristic timescales between hours and days, such as pulsating, eclipsing and exploding stars. This survey represents a unique laboratory to explore large etendue observations from cadences of about 0.1 days and to test new computational tools for the analysis of large data. This work follows a fully \textit{Data Science} approach: from the raw data to the analysis and classification of variable sources. We compile a catalog of ${\sim}15$ million object detections and a catalog of ${\sim}2.5$ million light--curves classified by variability. The typical depth of the survey is $24.2$, $24.3$, $24.1$ and $23.8$ in $u$, $g$, $r$ and $i$ bands, respectively. We classified all point--like non--moving sources by first extracting features from their light--curves and then applying a Random Forest classifier. For the classification, we used a training set constructed using a combination of cross-matched catalogs, visual inspection, transfer/active learning and data augmentation. The classification model consists of several Random Forest classifiers organized in a hierarchical scheme. The classifier accuracy estimated on a test set is approximately $97\%$. In the unlabeled data, $3\,485$ sources were classified as variables, of which $1\,321$ were classified as periodic. Among the periodic classes we discovered with high confidence, 1 $\delta$--scutti, 39 eclipsing binaries, 48 rotational variables and 90 RR--Lyrae and for the non--periodic classes we discovered 1 cataclysmic variables, 630 QSO, and 1 supernova candidates. The first data release can be accessed in the project archive of HiTS\footnote{\url{http://astro.cmm.uchile.cl/HiTS/}}.

\end{abstract}



\section{Introduction} \label{sec:intro}

Astronomy has entered the Time Domain era with large surveys that monitor the sky for several years aiming to study time variations of astronomical sources. Surveys like MACHO \citep{machoS93}, OGLE \citep{ogleS92}, EROS \citep{erosS93} and ASAS \citep{asasS97} have been used to discover new classes of variable sources and reach a better understanding of their properties and physical nature. These projects were designed with a specific scientific goal, such as the study of microlensing events as a signature of Massive Compact Halo Objects, which could explain Dark Matter, as in the case of MACHO \citep{machoV95}, OGLE \citep{ogleV94} and EROS \citep{erosV95}. Not only microlensing events were discovered but a large sample of RR-Lyrae, Eclipsing Binaries, Cepheids, Long Period Variables among others were also discovered. This led to a better calibration of period-luminosity relations and therefore more accurate distance estimators in the local Universe, e.g. \cite{Sesar17}.

Presently, there are surveys that are reaching deeper observations, searching for variable stars in the outer Halo of our Galaxy and in dwarf satellite galaxies of the Milky Way (MW). All of this thanks to new astronomical facilities: from medium to large sized telescopes, equipped with large field of view instruments like the Dark Energy Camera \cite[DECam, ][]{DECam15} at the 4m Blanco telescope. The key quantity that measures the survey capabilities of a telescope is etendue, which is the product of mirror collecting area and field of view. Current surveys are focused on transients and variables in general, like the Catalina Surveys \citep{Drake09}, the La Silla-Quest Variable Survey \citep{laSilla12}, the QUEST--La Silla AGN variability survey \citep{Cartier15}, Intermediate/Palomar Transient Factory \cite[PTF/iPTF, ][]{ptf09}, SkyMapper Southern Sky Survey \citep{SkyMapper07}, The Panoramic Survey Telescope and Rapid Response System \citep[PanSTARRS, ][]{PS16}, Korean Microlensing Telescope Network \citep[KMTNet, ][]{KMTNET16}, Gaia \citep{gaia16a}, the Hyper Suprime-Cam Subaru Strategic Survey \citep{Aihara17}, the Vista Variable in the Via Lactea (VVV, \citealt{VVV10}), and others. All these current projects are delivering huge amounts of raw data that need to be processed taking advantage of the available computational resources: High Performance Computing (HPC) to store, manage and analyze data of the order of terabytes; data visualization tools to analyze high dimensional data; Machine Learning algorithms to perform automatic classifications; and more.
These surveys are laboratories for the next generation of Time Domain instruments, such as the Large Synoptic Survey Telescope \cite[LSST, ][]{lsst09}, the Zwicky Transient Factory \cite[ZTF, ][]{Bellm14}. The data streams will scale exponentially, approximately following Moore's law (data volume doubles approximately every two years), therefore a more Data Science driven methodology must be implemented to extract the astrophysical knowledge. In particular, the combination of high cadence and large etendue observations represent a perfect combination to search for transients and variable sources with short characteristic time scales such as supernovae, RR-Lyrae, nearby asteroids, close eclipsing binaries and new types of transients.

Machine Learning (ML) methods have demonstrated their capability to provide solutions to various astronomical challenges. Important tasks such as classification of galaxies and stars using either photometric data (\citealt{Ball06}, \citealt{Vasconcellos11}, \citealt{Kim15}) or images (\citealt{Kim17}), classification of variable sources from time series data (\citealt{Richards12}, \citealt{Pichara13}, \citealt{Pichara16}, \citealt{Lochner16}), regression and model fitting algorithms (\citealt{Ball07}, \citealt{Cavuoti15}, \citealt{DIsanto18}), classification of transient events (\citealt{Mahabal08}, \citealt{Bloom12}, \citealt{Buisson15}, \citealt{Forster16}, \citealt{Wagstaff16}, \citealt{Cabrera17}), detection of anomalous events like new variability classes (\citealt{Nun16}) or unknown spectral signatures (\citealt{Baron17}, \citealt{Solarz17}, \citealt{Reis18}), represent some important examples. One of the advantages of using these methods compared to traditional methods of classification, is when data complexity is high. ML algorithms naturally deal with complex scenarios, not only in terms of the data volume, but also in the dimensionality of the problem (e.g., size of the feature space or resolution of the data). Nevertheless, not everything is straightforward when ML algorithms are used. One of the more common problems happens when supervised learning is applied. In the supervised approach, a model is trained with previously labeled data, and then the model is applied to new unlabeled data. Therefore a key aspect of ML is having a representative training set of the problem. This becomes difficult when new regimes of the parameter space are explored. For instance, in the analysis of time series from photometric data, this could correspond to deeper observations, new filters or the search for a new type of transient. Thus, one of the main challenges for ML applied in the astronomical domain is creating these training sets to be used with upcoming data.

In this article, we present two science products: a public catalog of detected point--like non--moving sources and an automatic classification of a catalog of variable stars in the HiTS database. HiTS was designed to find and study the early phases of supernova events using DECam imager at the 4--m Blanco Telescope on Cerro Tololo Interamerican Observatory (CTIO). We explain the source extraction and calibration process using standard calibrations and a comparison with public catalogs. We also present a statistical analysis of these catalogs and their structure. Finally, we follow a Machine Learning approach to automatically classify light curves according to their variability. In this final part we address the difficult task of constructing training sets. We explain the different strategies that were used following standard astronomical procedures like catalog cross--matching and visual inspection, modern ML techniques like Active Learning (AL) and Transfer Learning (TL) and data augmentation techniques based on transformations to the existing data in the labeled set.

The structure of this paper is the following: in Section \ref{sec:obs} the survey is presented; in Section \ref{sec:data_red} the pipeline structure is described: pre--processing, catalog creation, astrometric and photometric calibrations, and database structure; in Section \ref{sec:classification} we perform an automatic classification of the detected objects using light--curve features, training set construction, and classifier validation and classification; in Section \ref{sec:results} the classification results are presented and we discuss the encounter biases; and finally in Section \ref{sec:summary} the conclusions of this work are presented.

\section{Observations} \label{sec:obs}

The HiTS survey \citep{Forster16} consists of three observational campaigns in 2013,  2014 and 2015, aiming to study the early phase of supernova explosions. Due to this goal, the survey was designed to have a relatively high cadence, large field of view and high limiting magnitude. This combination of high cadence and large etendue offers a unique opportunity to do science other than supernova studies, such as studies of moving objects (\citealt{Pena18}), studies of distant RR-Lyrae stars (\citealt{Medina17}, \citealt{Medina18}) and variability studies in general.

In this work, we analyze data from all the 2013, 2014 and 2015 observational campaigns. In 2013, we observed 120 deg$^2$ in 40 fields during four consecutive nights, using an exposure time of 173 seconds, four times per night with a cadence of two hours  in the $u$ band. In 2014, we observed 120 deg$^2$ in 40 fields during five consecutive nights, using an exposure time of 160 seconds, four times per night with a cadence of two hours in the $g$ band. In 2015, we observed 150 deg$^2$ in 50 fields (with an overlap of 14 fields with 2014) during six consecutive nights, using an exposure time of 86 sec, five times per night with a cadence of 1.6 hours, mostly in the $g$ band, but also including $r$ and $i$ band observations. In 2014 and 2015, we also had imaging follow--up observations after the main run following an approximately logarithmic spacing in time. For more details of the survey see \cite{Forster16}.

\section{Data reduction} \label{sec:data_red}

The pre--processing was performed using a modified version of the DECam community pipeline \cite[DCP, ][]{Valdes14}, which performs electronic bias calibrations, crosstalk corrections, saturation masking, bad pixel masking and interpolation, bias calibration, linearity correction, flat field gain calibration, fringe pattern subtraction, bleed trail, edge bleed masking and interpolation. Cosmic rays were removed using a modified parallel version of CRBLASTER \citep{Mighell10}, which uses a Laplacian filter \citep{vanDokkum01}. We removed the DECam CCDs with known issues N30, S30 and S7 in all the fields\footnote{\url{http://www.ctio.noao.edu/noao/node/2630}}. The main difference between the data reduction process in this work compared to \cite{Forster16} is the use of images from individual epochs instead of subtracted images.

\subsection{Source Extraction} \label{subsec:SEx}
Detection and extraction of sources were performed using Source Extractor \cite[SExtractor][]{Bertin10} due to its speed, ease of implementation and ubiquitous use. A fine tuning of the parameters was done in order to go as deep as possible while keeping a false positive rate of detections in single frame images of less than 1.5\% and to perform well in point--like sources. This was done by comparing the single frame catalogs with those built from deeper stacked images as explained in Section \ref{subsec:depth}. We report fixed--aperture photometry using aperture diameter equal to multiples of the Image Quality (IQ), defined as half of the empirical full--width--half--maximum (FWHM) of the image, as well as Kron aperture photometry for extended sources, although SExtractor parameters were not optimized for this purpose. The SExtractor configuration parameters are presented in Table \ref{tab:SEx}. 

\begin{deluxetable}{cc}
\tablecaption{SExtractor input parameters \label{tab:SEx}}
\tablehead{\colhead{Parameter} & \colhead{Value}
}
\startdata
ANALYSIS\_THRESH	& 1.5 \\
BACK\_SIZE			& 64 [pix] \\
DETECT\_MINAREA		& 3 [pix] \\
DETECT\_THRESH		& 1.3 \\
GAIN				& 4.025 [electrons/adu]\tablenotemark{a} \\
PHOT\_AUTOPARAMS	& [2.5,3.5] \\
PIXEL\_SCALE		& 0.27 [arcsec/pixel]\tablenotemark{a} \\
SATUR\_LEVEL		& 44\,144 [adu]\tablenotemark{a} \\
WEIGHT\_GAIN		& YES \\
WEIGHT\_TYPE		& MAP\_VAP\tablenotemark{b} \\
\enddata
\tablenotetext{a}{Read from image header.}
\tablenotetext{b}{If available, if not WEIGHT\_GAIN is used.}
\end{deluxetable}

\subsection{Astrometric calibrations} \label{subsec:astrocal}

Output catalogs from SExtractor were astrometrically calibrated against the GAIA DR1 catalog \citep{gaia16b} using the latest version of the Software for Calibrating Astrometry and Photometry \cite[SCAMP v2.6, ][]{SCAMP}. The last version of SCAMP takes into account the full mosaic image from DECam to achieve the astrometric solution. As a comparison we perform cross--matching with GAIA and PanSTARRS Data Release 1 (PS1) catalogs. The root--mean--square of the residuals is in the order of $0.08$ and $0.05$ arcsec, respectively, with $99 \%$ of the cross--matched sources within $0.5$ arcsec of their matching object.

\subsection{Photometric calibrations} \label{subsec:photocal}

We adopt two different photometric calibration strategies depending on whether there is information from the PS1 reference catalog in the same filter of our observations. If available, we calibrate all epochs against PS1. If not available, we calibrate all epochs relative to a reference epoch with good observational conditions which are assumed to be photometric for computing zero points. We will call these calibration strategies C1 and C2, respectively. 

The problem of photometric calibrations can be considered from the point of view of their relative and absolute calibration accuracy. We have found that in those fields where PS1 is available, C1 or C2 give the same quality of relative photometry. However, C1 gives significantly better absolute photometric calibrations than C2. For 2014 and 2015 we calibrate all epochs using C1, whereas for 2013 we use C2. In what follows both strategies are described in details.

\paragraph{{\bf C1:}}
The absolute calibration against PS1 was performed by fitting  zero-points (ZP) for every field, CCD, band, photometry type (Kron and aperture), and epoch. We calculate the ZPs on cross-matched objects between the PS1 and HiTS catalogs with magnitudes between 16 and 21 for the $g$ band, and between 16 and 20 for the $r$ and $i$ bands. This is due to the difference in depths and observing conditions encountered when observing each band, namely airmasses or sky brightness. We applied a sigma--clipping filter of three times the median absolute deviation around the median of the magnitude difference distribution, to remove outliers. ZP values were calculated as the median value of the magnitude difference for sources that remained after filtering. We estimated uncertainties on ZPs using bootstraping to sample from the distribution of filtered magnitude differences. We tested these results against a Monte Carlo Markov Chain (MCMC) method assuming a model with the ZP and its error as parameters. The MCMC posterior median error was in average 16\% larger than the bootstrap error. Thus, we report ZP uncertainties as $1.16$ times the bootstrap error. Finally, reported magnitudes for detected sources are corrected by the ZP calculated as:
\begin{equation} \label{eq: PS_mag}
m = -2.5\log (F) + 2.5\log (t_{exp}) + ZP,
\end{equation}
\begin{equation} \label{eq: PS_err}
\delta m = \sqrt{\left(\frac{2.5 \delta F}{F\ln(10)}\right)^2 + (\delta ZP)^2},
\end{equation}
where $m$ is the calculated AB--magnitude, $F$ is the SExtractor measured flux in analog--to--digital--units (ADU), $t_{exp}$ is the exposure time of the observation, $ZP$ is the zero--point mentioned above, $\delta m$ is the calculated photometric uncertainty, $\delta F$ is the measure flux error (from Equation \ref{eq: pix_corr} as explained next) in ADU and $\delta ZP$ is the uncertainty of the ZP.
\paragraph{{\bf C2:}}
We applied relative flux calibrations with respect to a reference epoch, which was chosen to be the best in terms of seeing for every field and CCD. We fitted a linear relation between the fluxes of the same stars in the non--reference vs reference epochs and applied a transformation to the fluxes in the non--reference epochs: $F'_i = F_i / a_{\rm flux}$, where $F'_i$ is the transformed flux, $F_i$ is the original flux, and $a_{\rm flux}$ is the slope of the fitted relation. Then, all fluxes are converted to magnitudes following the DECam Photometric Standard Calibration, ignoring color terms since the relevant colors are not available.
\begin{equation} \label{eq: DECam_mag}
m = -2.5\log (F') + 2.5\log (t_{exp}) - A - K X,
\end{equation}
\begin{equation} \label{eq: DECam_err}
\delta m = \sqrt{\left(\frac{2.5 \delta F'}{F'\ln(10)}\right)^2 + (\delta A)^2 + (X\delta K)^2},
\end{equation}
where $F'$ is the transformed flux mentioned above in ADU, $X$ is the observed airmass, $A$ and $K$ are the zero--point and airmass coefficients from the DES science verification archive \footnote{\url{https://cdcvs.fnal.gov/redmine/projects/des-sci-verification/wiki/Standard_Star_Photometry}}, $\delta F'$ is the transformed flux error in ADU, $\delta A$ and $\delta K$ are the reported uncertainty of the zero--point and airmass coefficients.

\begin{figure}
\center
\includegraphics[width=.47\textwidth]{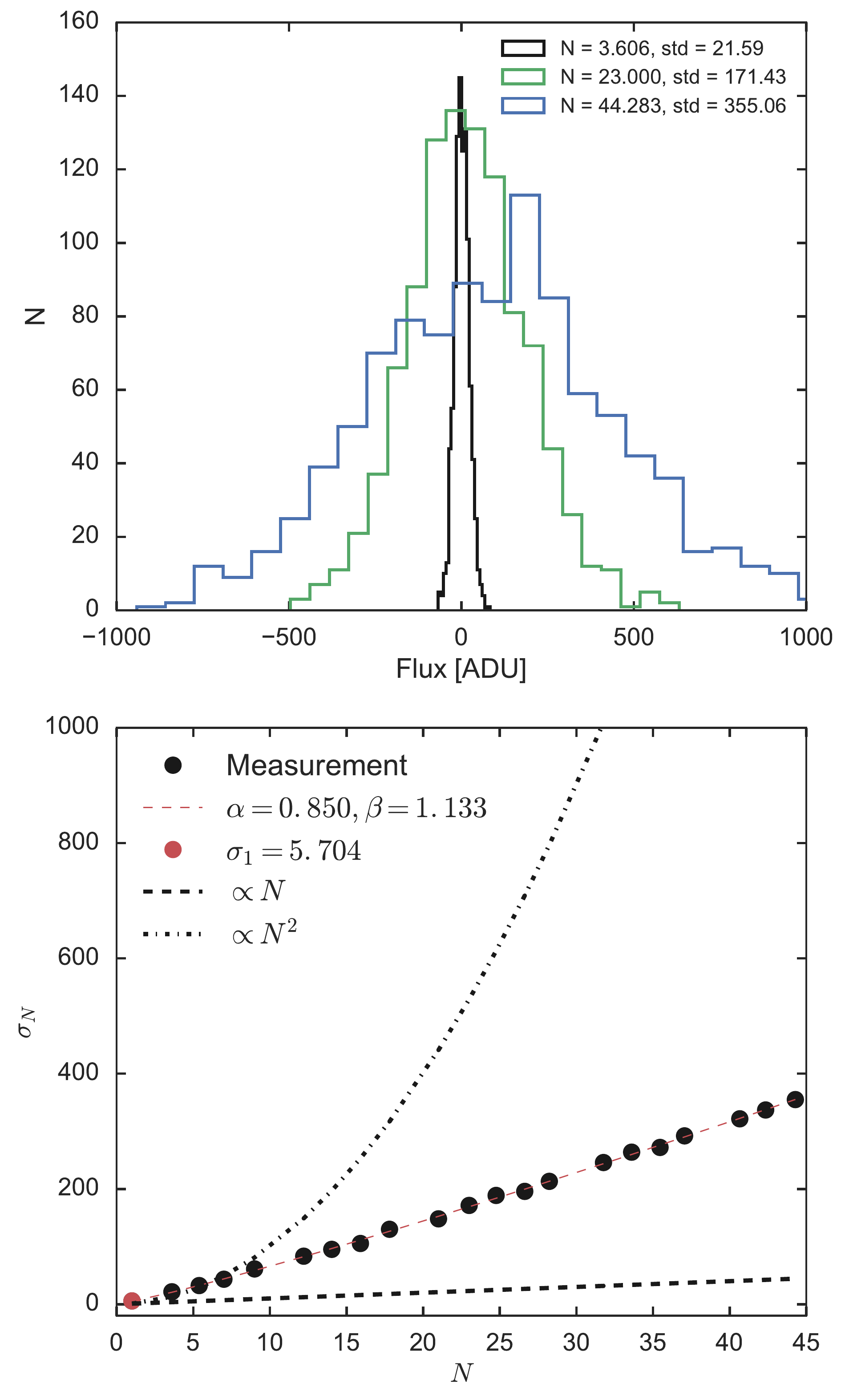}
\caption{Top: distribution of measured flux on randomly located apertures for three effective size $N$, the fluxes were measured on the background subtracted image. Bottom: background fluctuations as a function of aperture size. Circles represent the empirical variation on images of field Blind15A\_25, CCD N1, red segmented line represents the fitted function with $\alpha = 0.850$ and $\beta = 1.133$. Both black dashed lines represent the extreme cases of independent noise from pixel to pixel and correlated noise from fluctuations in background level which leads to a rms proportional to $N$ and $N^2$, respectively. \label{figure:skyfluc}}
\end{figure}

In order to estimate the uncertainties $\delta F$ of the measured fluxes we use corrected SExtractor errors. It is well known that errors from SExtractor are generally underestimated (\citealt{Labbe03}, \citealt{Gawiser06}). SExtractor estimates errors using the following relation \citep{Bertin10}:
\begin{equation} \label{eq: sex_err}
\delta F^{2} = \sigma^{2}_1 n_{pix} + \frac{F}{GAIN},
\end{equation}
where $\sigma_1$ is the typical fluctuation per pixel (mostly due to sky Poisson noise), $n_{pix}$ is the number of pixels within the aperture, $F$ is the measured flux and $GAIN$ is the effective gain used to convert ADUs into detected photons. The first term represents the sky fluctuations assuming uncorrelated noise between pixels and the second term is the Poisson variance or shot--noise from the source. One way to include the noise correlation is to empirically study background fluctuations as a function of the aperture size using  randomly located circular apertures with a range of aperture diameters \citep{Gawiser06}. In Figure~\ref{figure:skyfluc}, top panel, we show the distribution of flux measurements of background subtracted images for 1000 randomly located apertures with different effective sizes $N$. Then, we model the uncertainties to be proportional to the number of pixels of the aperture to a given power $\beta$:
\begin{equation} \label{eq: emp_rel}
\sigma_N = \sigma_1 \alpha N^{\beta},
\end{equation}
where $N$ is the effective size of the aperture ($N = \sqrt{n_{pix}}$). In Figure \ref{figure:skyfluc}, bottom panel, we show the empirical relation (circles) and the extreme cases of no correlation and perfect correlation which correspond to ${\sim} N$ and ${\sim} N^2$, respectively.

Finally, the uncertainties of the measured fluxes are given by:
\begin{equation} \label{eq: pix_corr}
\delta F^{2} = \sigma^{2}_1 \alpha^2 n^{\beta}_{pix} + \frac{F}{GAIN},
\end{equation}
\noindent were $\alpha$ and $\beta$ are given in equation~\ref{eq: emp_rel} and $n_{pix}$ is the number of pixels within the aperture. For Kron aperture on SExtractor, the area is given by the Kron best--fit ellipse. Then the number of pixels within the Kron aperture is given by: $n_{pix} = \pi r^{2}_{kron} \times A\_IMAGE \times B\_IMAGE$, where $r_{Kron}$ is the Kron radius and $A\_IMAGE$ and $B\_IMAGE$ are the major and minor axis of the ellipse, respectively. For fixed circular apertures the number of pixels is given by $n_{pix} = \pi \times (k \times IQ)^2$ where k is an integer.

\subsection{Survey depth} \label{subsec:depth}
\begin{figure}[ht!]
\includegraphics[width=.45\textwidth]{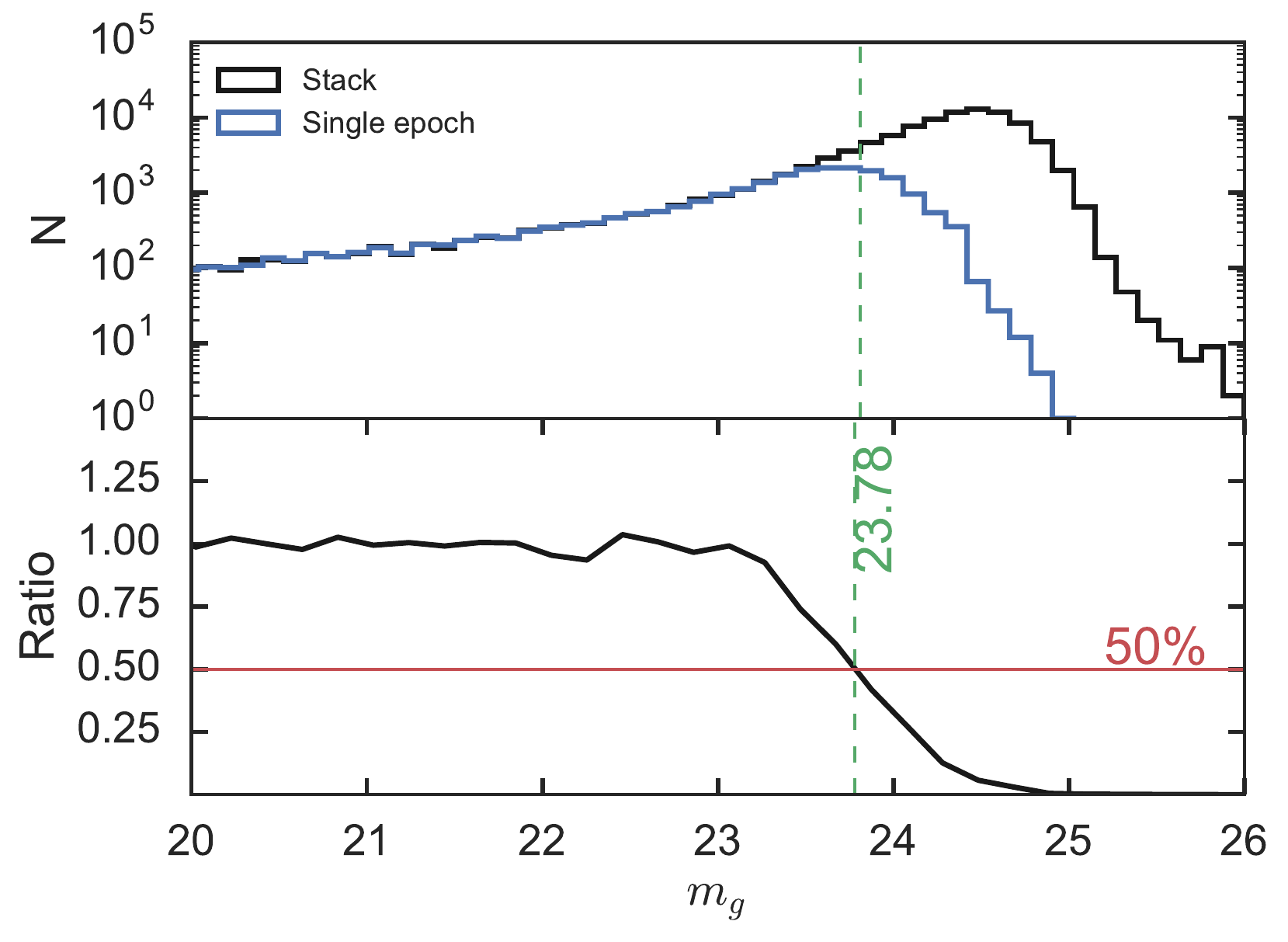}
\caption{\textit{Top}: distribution of detected sources, black histogram represents all detection in stack image and blue histogram are sources from single epoch catalog with a match between both. \textit{Bottom}: ratio of both distribution representing the detection recovery, horizontal red line represents the $50\%$ level of completeness, this is achieved at ${\sim} 23.8$ magnitude for a typical field in $g$ band. \label{fig:complet}}
\end{figure}

In order to calculate the \emph{completeness magnitude} of our catalogs, we created deep images combining the 10 best epochs per campaign. Catalogs were extracted from stacked images and then compared against catalogs from single epoch images. Assuming that at the depth of the single images the stacked images were complete (in terms of detectable sources), we compared the magnitude distribution from both catalogs to obtain the magnitude at a given completeness level.
In the upper panel from Figure \ref{fig:complet} the black and blue lines represent the distribution of detected sources in the stack and single epoch images, respectively. It is easy to see that at the bright end (left) the ratio is consistent with 1, see lower panel, but around 23th magnitude this ratio decreases, reaching $50\%$ at 23.8 magnitudes.

In order to calculate the \emph{limiting magnitudes} per field and epoch we measured flux fluctuations of randomly located empty apertures, using the standard deviation of 1000 aperture fluxes where no sources were detected. Then, assuming a signal--to--noise--ratio ($S/N$) of 5 for detection we determined the limiting magnitudes per field, CCD and epoch as 5 times the previous standard deviations.

\begin{figure*}[ht!]
\gridline{\fig{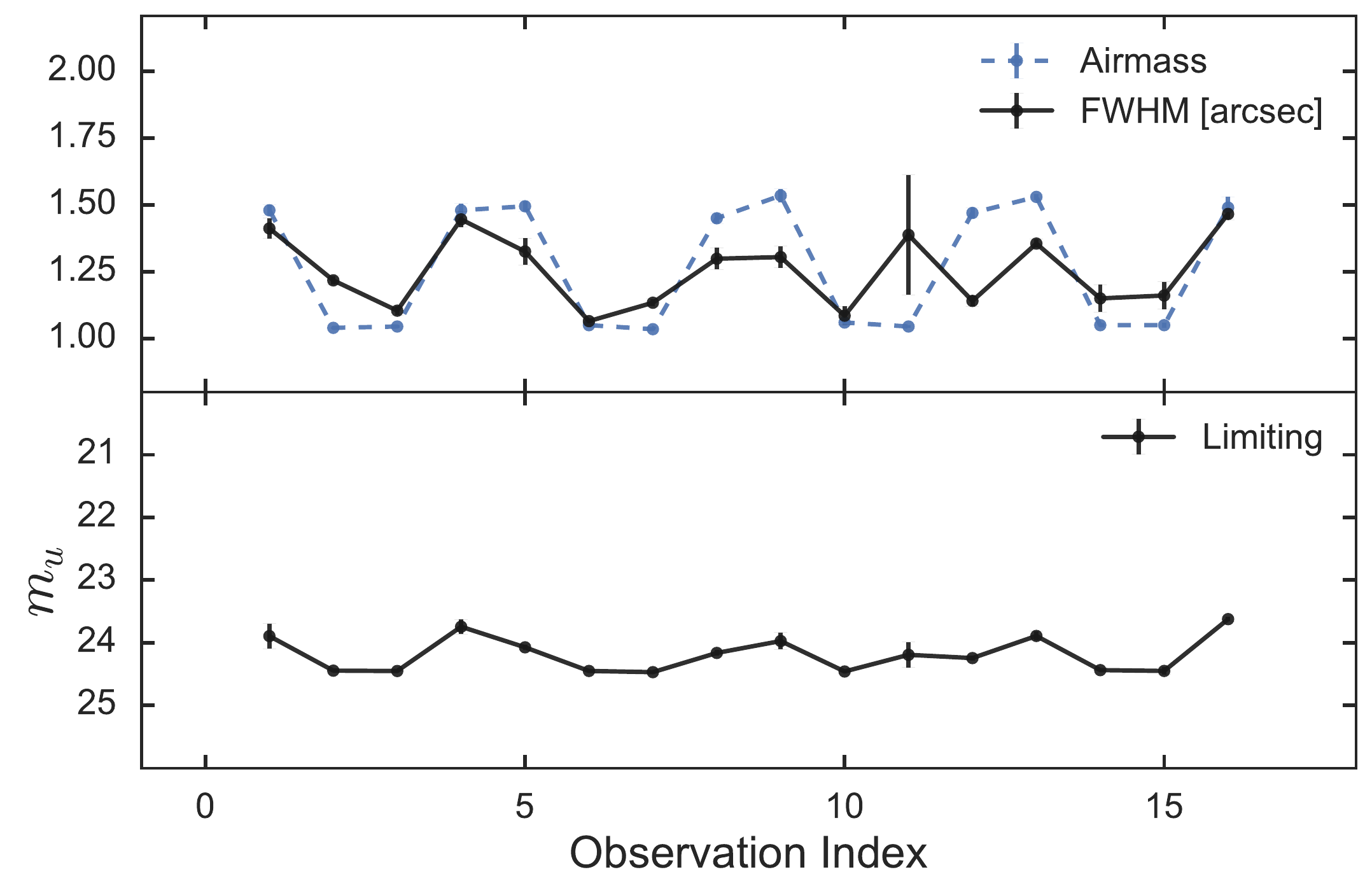}{0.33\textwidth}{(a)}
          \fig{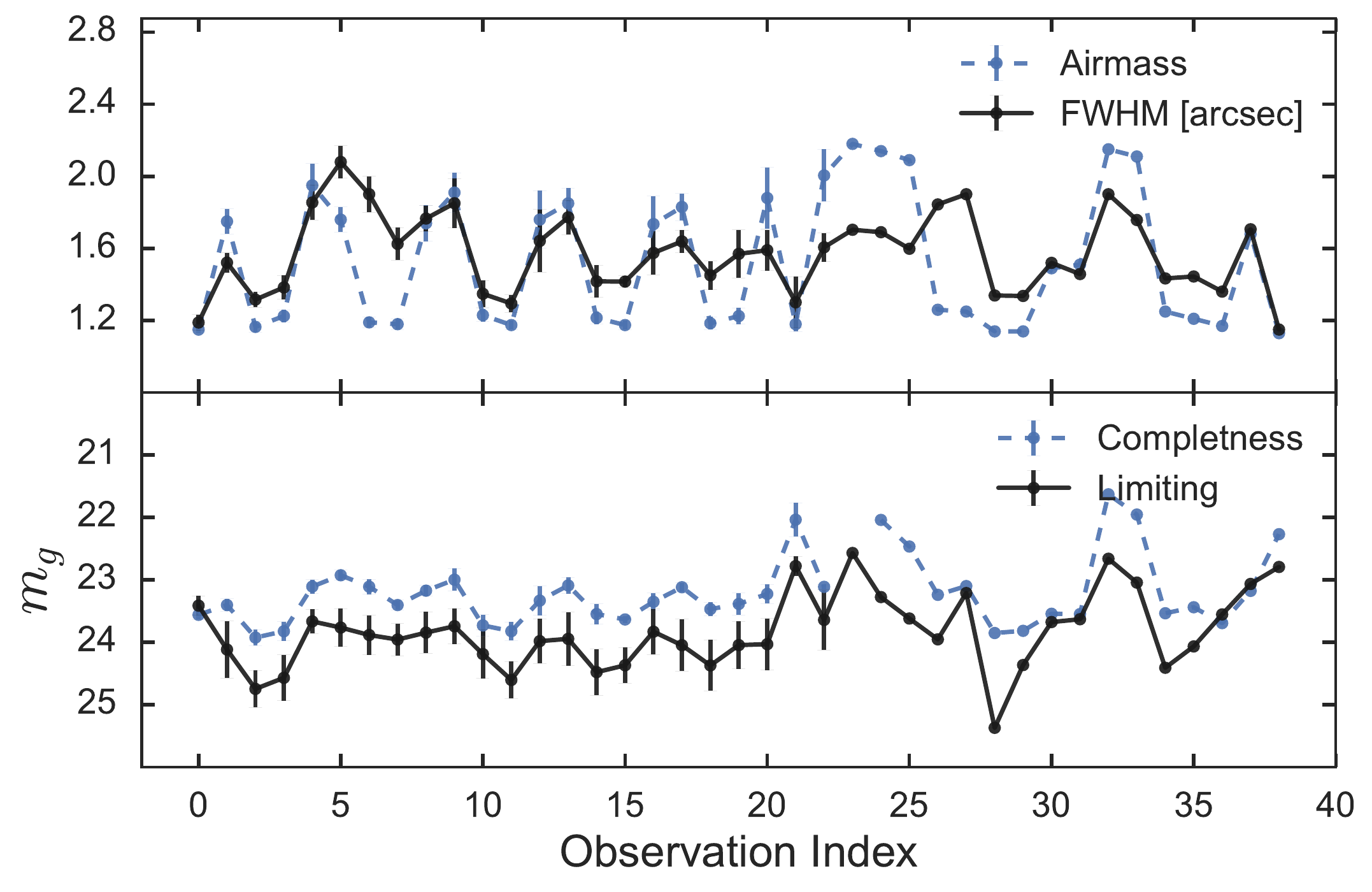}{0.33\textwidth}{(b)}
          \fig{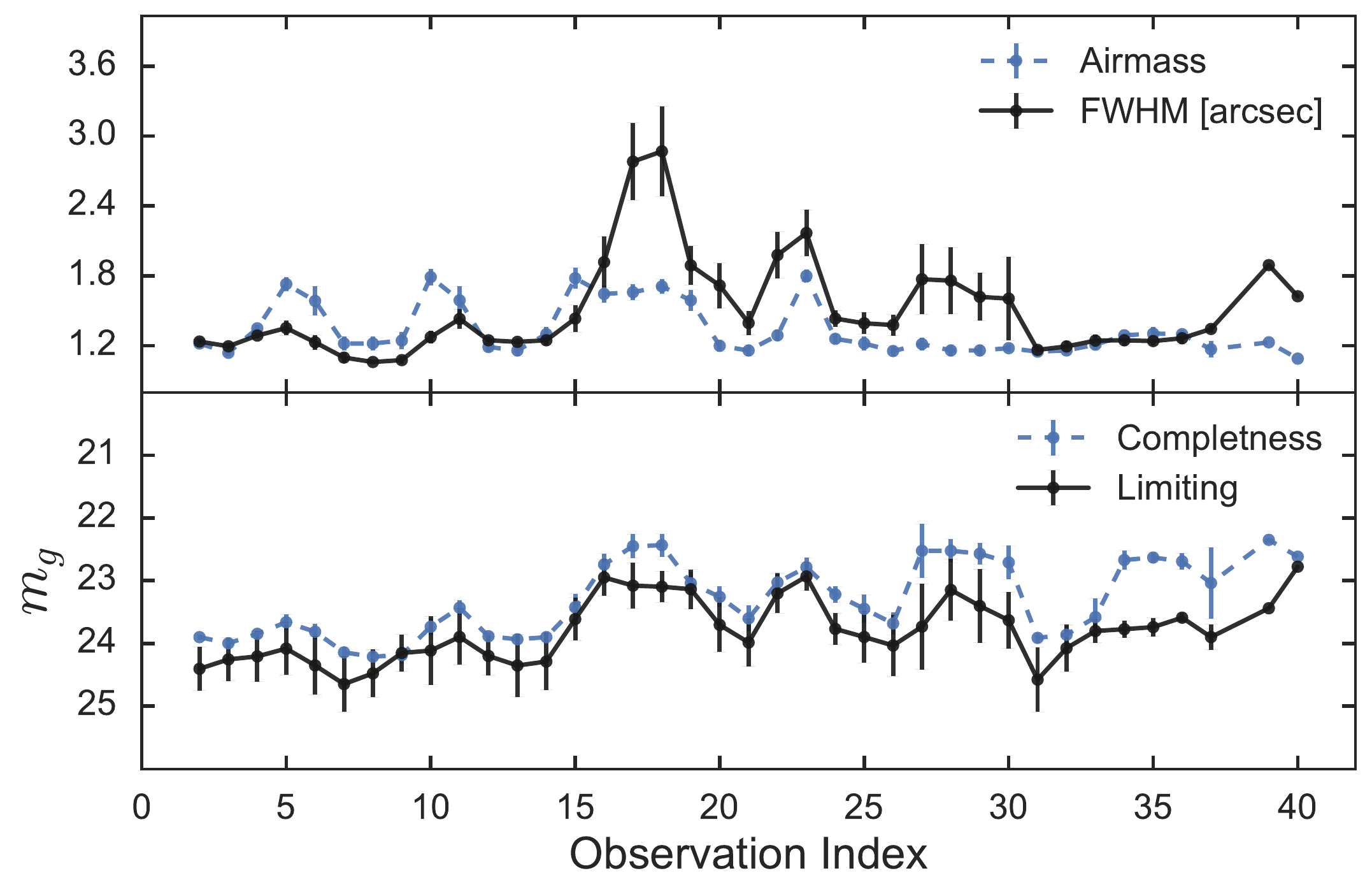}{0.33\textwidth}{(c)}
          }
\caption{Panels (a), (b), and (c) refer to observation campaigns during 2013, 2014 and 2015, respectively. Evolution of the observed airmass (dashed blue line) and measured FWHM (solid black) in arcseconds (\textit{top}). Derived completeness and limiting magnitude are shown in the \textit{bottom} panel, both as a function of epoch number.
\label{fig:limits_HiTS}}
\end{figure*}

Figure ~\ref{fig:limits_HiTS} shows in the lower panels the evolution along epochs of g band \emph{limiting} (solid black line) and \emph{completeness} (dashed blue line) magnitudes for observations during 2013, 2014 and 2015, respectively. Upper panels show evolution of observed airmass and measured FWHM in arcseconds. In all three years, the \emph{limiting} and \emph{completeness} magnitudes follow the airmass evolution within the night. Figure~\ref{fig:limits_HiTS} (c) panel, shows an increase of the measured FWHM between epochs 15 and 25, this is due to observations during two cloudy nights, having a severe impact in the limiting magnitudes.

The rate of detections that are non--real sources, i.e. the False Positive Rate of detections, were also derived comparing catalogs from stacked images and single epoch images, which were found to be at the level of $1\%$ at magnitude $23.8$ (only considering point-like sources). Therefore, our catalogs are typically at least 99\% pure at the completeness magnitude.

\begin{figure}[ht!]
\includegraphics[width=.47\textwidth]{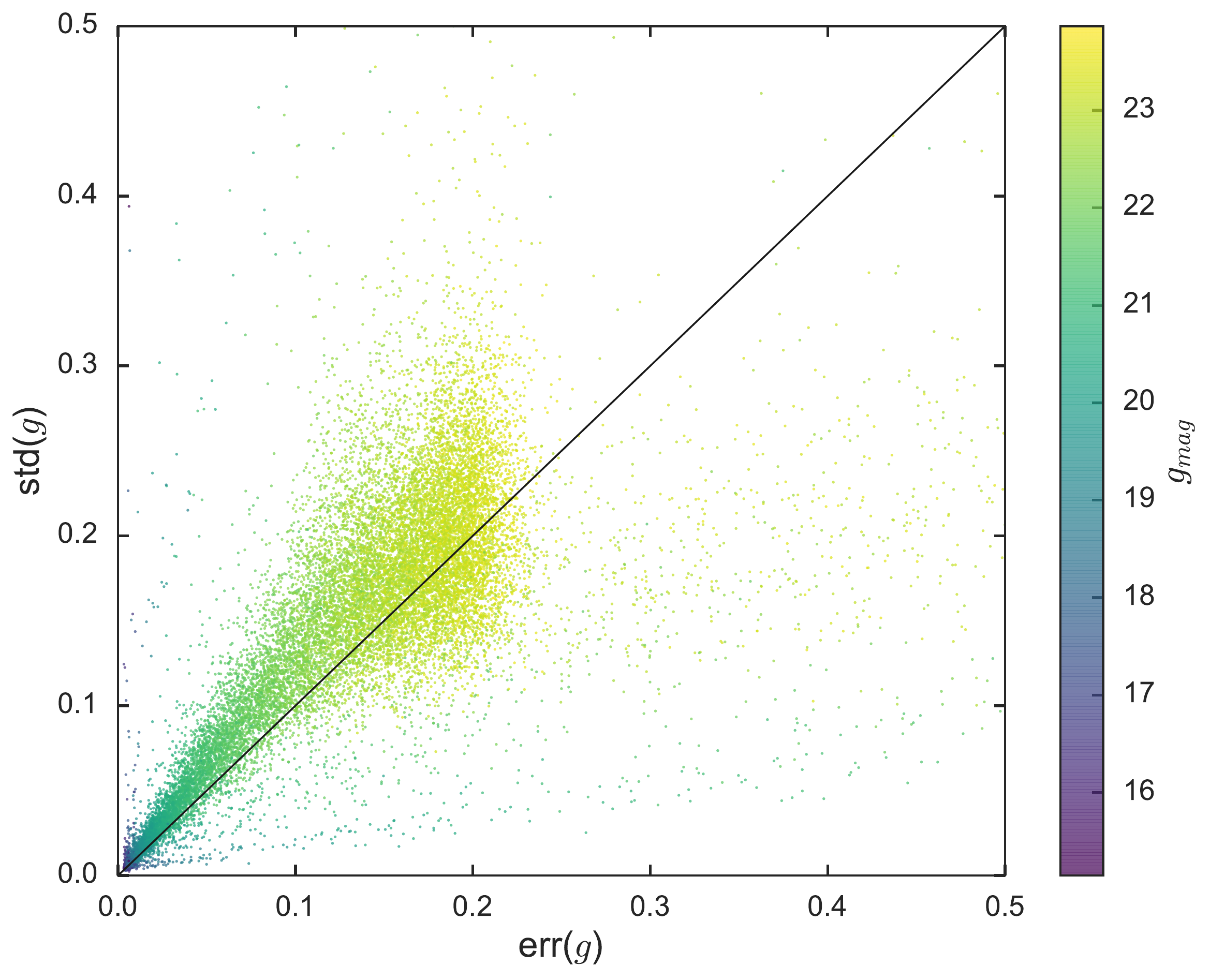}
\caption{Standard deviation of detections vs median photometric uncertainty color coded by median magnitude for a typical field during 2015 $g$ band. The diagonal line is the identity representing variations in the photometry similar to its error, significant departures from the identity are due to intrinsic variability and insufficient correction by pixel correlations. \label{fig:errors_15A_g}}
\end{figure}

In order to test the quality of our photometric errors, we compared the flux standard deviation with the median photometric uncertainty for every source. In Figure \ref{fig:errors_15A_g}, we show the empirical standard deviation vs the median estimated photometric uncertainty. We found that the previous standard deviations are of the same order as the estimated median errors (combining SExtractor, pixel correlation and zero--point uncertainty). However, the distribution of the ratio between the previous two quantities has a median of about $1.3$. An increased empirical standard deviation could be due to several factors: the contribution of the more noisy epochs due to varying observational conditions; the correction in Equation \ref{eq: pix_corr}, which does not take into account correlated noise coming from the source; and the contribution of intrinsically variable sources. Thus, the ratio between empirical standard deviations and the estimated uncertainties are expected to be greater than 1.

\begin{figure}[ht!]
\includegraphics[width=.47\textwidth]{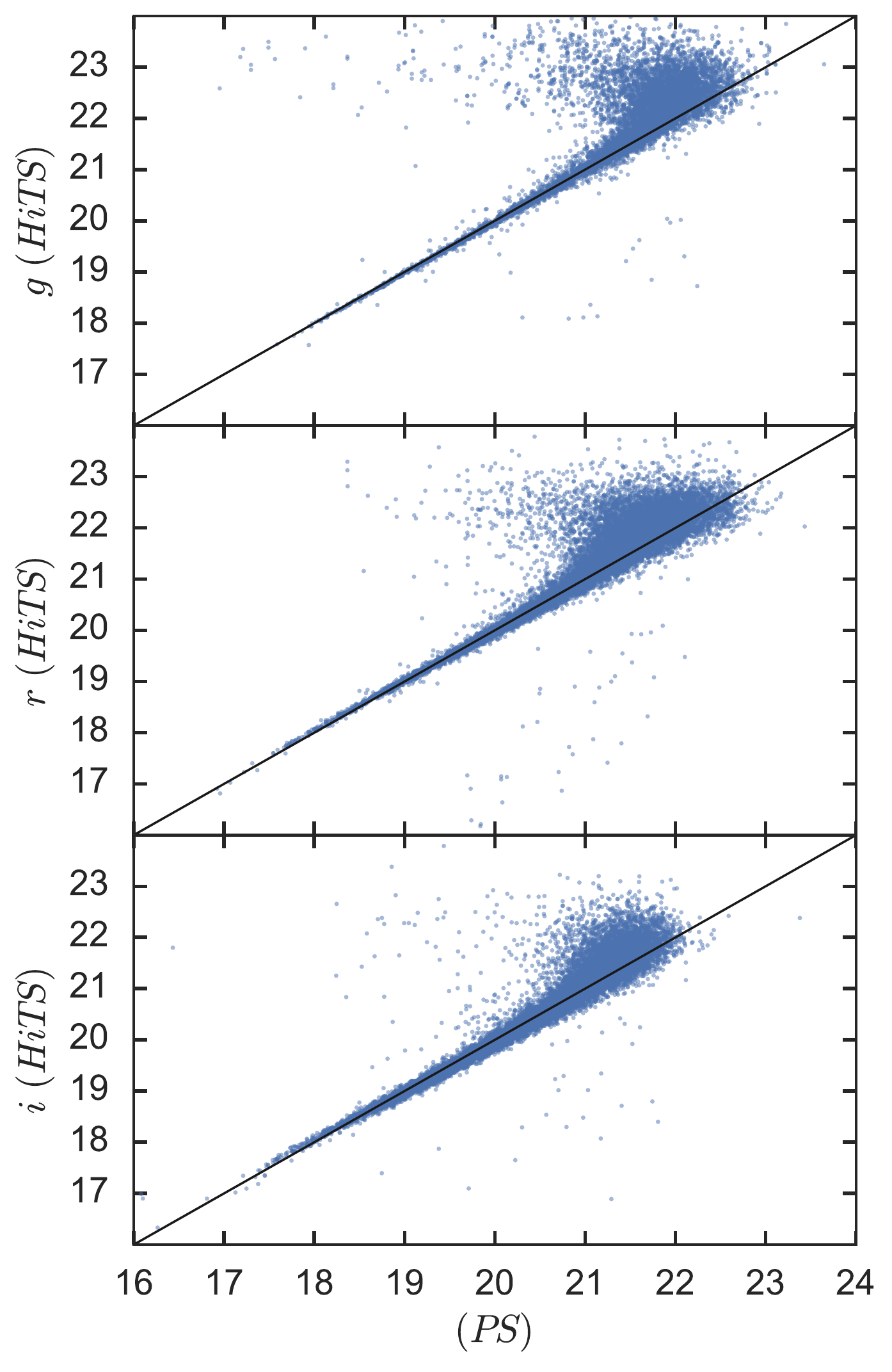}
\caption{HiTS against PS1 photometry for $g$, $r$ and $i$ bands in the top, middle and lower panels, respectively. The scatter around the identity for relatively bright sources, up to 21.5, is of the order of $0.02$, $0.04$ and $0.07$ for $g$, $r$ and $i$ bands. Beyond 21.5 magnitudes the scatter from the identity increase, we claim that PS1 values tend to be underestimated compared to HiTS photometry mainly due to the single epoch depth of both surveys, where HiTS reach the completeness magnitude at ${\sim}24$ and PS1 at ${\sim}23$. \label{fig:HiTS_vs_PS1_15A_g}}
\end{figure}

Further comparisons of HiTS against PS1 photometric catalogs are shown in Figure \ref{fig:HiTS_vs_PS1_15A_g} for all three filters available in the 2015 data. In $g$ band the relation is close to the identity with a small scatter of the order of $0.02$ mag. up to 21.5 magnitudes. Above 21.5 mag, where the HiTS catalog is three magnitudes and PS1 two magnitudes below its completeness magnitude, PS1 values tend to be underestimated compared to the HiTS photometry. For $r$ and $i$ band the conclusions are similar, but the scatter is of the order of $0.04$ and $0.07$ mag., respectively. 

\subsection{Catalogs and Database} \label{subsec:catalog}
In total, we obtained $1\,980\,107$ detected sources in the \textit{u} band for 2013, $5\,389\,028$ sources in the \textit{g} band for 2014. For 2015 observations, we obtained $5\,117\,233$ sources in the \textit{g} band, $5\,884\,126$ the \textit{r} band and $4\,572\,003$ in the \textit{i} band. Due to the overlap of 14 fields between 2014 and 2015 observation campaigns, $1\,190\,008$ sources have been cross--matched using a radius of 0.5 arcsec between both catalogs. The number of single epoch detections for all three campaigns is about $100$ millions. The structure of this catalog is similar to the structure used in PS1. The column description of the catalogs is presented in Appendix Section \ref{apx: cat_ex} (see Table \ref{table:catalog_cols}). Public catalogs are available and accessible in the project archive of HiTS\footnote{\label{dr1}\url{http://astro.cmm.uchile.cl/HiTS/}}. The database is available as compressed tarball files for the entire HiTS survey and also separated by fields. We will test in the future the possibility to use a dedicated time series database (e.g. influxDB\footnote{\url{https://www.influxdata.com/}}) to store and access the data. 

\section{Automatic Classification} \label{sec:classification}
The typical procedure to perform a supervised automatic classification using Machine Learning algorithms is the following: first build a labeled set with the desired classes in a subset of the entire database, this can be accomplished by running cross--match with public databases available; then train a supervised classifier with a subset of the labeled set, the training set; then the model is validated with a subset of the labeled set which was not used to train, the test set; and finally a prediction is made on the unlabeled data. In our case we use a feature engineering approach, i.e. to represent each time series as a user--defined feature vector, and with labels corresponding to variability classes of the astronomical sources.

We clean the detection catalog to remove extended sources, we filter by $FWHM,\  Ellipticity,\ FluxRadius$ and $KronRadius$ to select point--like sources. Afterwards, we filtered out saturated sources at the bright end of the magnitude distribution. Finally, we select all sources with more than fifteen detections to build their light curves and perform the automatic classification by variability. We end up with $2\,536\,100$ point--like sources after the selection mentioned above for the 2014 and 2015 data. 

\subsection{Feature extraction} \label{subsec:features}
We extract features using the Feature Analysis for Time Series \citep[FATS, ][]{Nun15}. Some of these features are: mean, standard deviation, amplitude, period of the Lomb-Scargle periodogram maximum, false alarm probability of this period, mean variance, median absolute deviation. The full list of features is shown in Table \ref{tab:feat_list} in the Appendix Section \ref{apx_sec: feature}. Note that the set of features has already been used in other works (\citealt{Nun16}, \citealt{Pichara16}, \citealt{Mackenzie16}). We also added periods calculated using both the Generalized Lomb-Scargle technique \cite[GLS, ][]{Zechmeister09} and Correntropy kernel periodogram \cite[CKP, ][]{Huijse12}, as well as color indices calculated when observations were available ($g$, $r$ and $i$ band).

\subsection{Labeled Set} \label{subsec:labeled}
In order to build a labeled subset with the astronomical variability classes expected to appear in our survey, we considered three approaches: cross--match with public catalogs, visual inspection, and TL combined with AL.

The first and the more standard  approach is to run a cross--match with public catalogs of variable sources. However, due to the uniqueness of our survey in terms of cadence, depth and survey area, other surveys tend to have a small overlap with HiTS. Surveys such as MACHO, EROS and OGLE do not overlap spatially with HiTS. We found overlap with the Sloan Digital Sky Survey Data Release 9 \citep[SDSS-DR9, ][]{SDSSDR9_BOSS}, the General Catalog of Variable Sources \citep[GCVS, ][]{GCVS}, the Catalina Sky Survey Data Release 1 \citep[CSDR1, ][]{CSDR1}, and The International Variable Star Index \citep[VSX, ][]{VSX}. We also included a comparison with parallel searches for RR Lyrae and supernovae on the same data \citep[][respectively]{Medina18,Forster16}. It is important to notice that the low number of positive cross--matched sources (about 30) between our light curves and the SNe detected by \citeauthor{Forster16} (about 120 detections) is due to the methodology used. We detect sources with SExtractor in the direct images, while \citeauthor{Forster16} used image differences which perform better near the core of the galaxy host. The classes and number of cross--matched sources from each different survey are summarized in Table~\ref{tab:ts_nums}.

The second approach consists of a visual inspection of light curves from sources that have a low false alarm probability obtained from their Lomb-Scargle periodogram maximum (i.e., the largest \textit{period\_fit} value given by FATS). We also visualized the light curves of sources with very low variability (\textit{std, mean variance} and \textit{variability index} given by FATS) in order to add them to the non-variables class of the training set.

After the first two approaches we end up with eleven classes of variability: non--variables (NV), Quasars (QSO), cataclysmic variables (CV), RR-Lyrae (RRLYR), eclipsing binaries (EB), miscellaneous variable stars (MISC), supernovae (SNe), long period variables (LPV), rotational variables (RotVar), ZZ Ceti variables (ZZ) and $\delta$--Scuti variable stars (DSCT).
 
Due to the small amount of instances for some classes after the cross-match, which leads to an unbalanced training set, we tried the third approach of TL combined with AL.
Transfer Learning \citep{TL} is a method to learn from a training dataset which exists in a different domain, the source domain, to train a classifier that will be applied in a different domain, the target domain. For instance, to train a classifier based on $V$-band light curves which will be applied to $g$-band light curves. This is particularly useful when a large training set exists in the source domain, but no equivalent set exists in the target domain.
Active learning \citep{AL} is an iterative and interactive process where expert input is required to classify objects where the classification cannot be done with a high level of confidence, and with this information improve the classifier. This process is done iteratively until certain criteria are met, e.g. the test set accuracy or the amount of sources per class are satisfactory.

There are already compiled training sets which can be used for the classification of variable stars, such as MACHO, OGLE and EROS. A transfer learning approach is required to train a classifier for HiTS using these datasets because these surveys were observed with a different band, cadence, depth, and in a different part of the sky (the source domain for TL). A simple way to transfer the labeled sources is to calculate the same features in both the source and target domain, and compare their distributions. Then a simple transformation is found, e.g. scaling and translation, which forces both distributions to be similar.
The main problem with this approach is that finding the transformation can be difficult, especially when the distributions of features are very dissimilar. This is our case. The sampling function of HiTS is very different to those of MACHO. MACHO was observed in $BVR$ filters, but HiTS was observed in the $ugri$ filters; MACHO reached a depth of ${\sim} 20$ mag in the $B$ band, while the HiTS depth was ${\sim} 24.5$ mag in the $g$ band; finally the cadence of MACHO is in the order of days (even weeks) during 5 years, but that of HiTS was in the order of hours during about a week. The latter is probably the most important difference between these surveys. Therefore, it was not possible to find a simple transformation between both spaces.

We tested a basic combination of TL with AL. A simple transformation was applied to the feature distribution of MACHO to match HiTS, and then a Random Forest classifier (see next subsection for a description of the method) was trained using the transformed feature values of MACHO. The hyper-parameters of the model were tunned following a cross--validation approach\footnote{Technique used to measure the model errors where the data are partitioned into train and test sets and the model is trained and tested several times, each one with a different combination of train/test sets. Finally, the error is computed from the statistics of all the trained models.\citep{astroMLText}.} within the MACHO training set. For this and further tests, we set a Stratified 6--Fold cross--validator to preserve class unbalance.
When classifiers without TL and with our simple transformation were tested on the current HiTS labeled dataset (using only labeled sources from cross--matching in Table \ref{tab:ts_nums}), we found that in the first case the accuracy was ${\sim} 5\%$, and in the second case the accuracy improved to ${\sim} 15\%$. None of these classifiers reached a satisfactory level of accuracy. But predicting in unlabeled HiTS data we were able to confirm the classification of those objects with the highest classification confidence via visual inspection done by experts, to then add them to the labeled set (basic AL procedure).
We performed this method with only one iteration, since most of the newly labeled objects belonged to the already well populated classes. In those cases the new objects were not added to the labeled set. EB sources provided by this method were later confirmed via post cross--matching with the same catalogs presented above. The contribution of this mixture of TL and AL to the training set is shown in Table \ref{tab:ts_nums}.


\floattable
\begin{deluxetable}{ccccccccccccc}
\tablecaption{Summary of number of objects per class from different approaches to building the labeled set. Variability classes are: non--variable (NV), quasar (QSO), cataclysmic variable (CV), RR--Lyrae (RRLYR), eclipsing binary (EB), miscellaneous variable (MISC), supernovae (SNe), long period variable (LPV), rotational variable (ROTVAR), ZZ Ceti variable (ZZ) and $\delta$--Scuti variable (DSCT). \label{tab:ts_nums}}
\tablehead{
\colhead{Source} & \colhead{NV} & \colhead{QSO} & \colhead{CV} & \colhead{RRLYR} & \colhead{EB} & \colhead{MISC} & \colhead{SNe} & \colhead{LPV} & \colhead{RotVar} & \colhead{ZZ}  & \colhead{DSCT} & \colhead{Total}
}
\startdata
SDSS-DR9		        & -    & 3\,495 & 85 & -   & -   & - & -  & -  & -  & -  & -  & 3\,580     \\
GCVS			        & -    & -    & 1  & 22  & 1   & - & -  & -  & -  & -  & -  & 24     \\
CSS 			        & -    & -    & 1  & 26  & 91  & - & -  & 1  & 5  & -  & -  & 124     \\
VSX 			        & -    & -    & 11 & 126 & 105 & 7 & -  & 1  & 5  & 2  & 1  & 258     \\
\citealt{Medina18} & -    & -    & -  & 60  & -   & - & -  & -  & -  & -  & -  & 60     \\
\citealt{Forster16}		& -    & -    & -  & -   & -   & - & 29 & -  & -  & -  & -  & 29    \\
Visual Inspection       & 5\,000& -    & -  & -   & -   & - & -  & -  & -  & -  & -  & 5\,000     \\
TL and AL  		        & -    & -    & -  & -   & 10\tablenotemark{a}& - & -  & -  & -  & -  & -  & 10 \\ \hline
Total\tablenotemark{b}  & 5\,000 & 3\,495 & 94 & 177 & 110 & 7 & 29 & 1  & 5  & 2  & 1  & 8\,921  \\ \hline
Data Augmentation       & 5\,000  & -      & -  & 100 & 200 & - & -  & - & 179 & - & 166 & 5\,645\\
Labeled set\tablenotemark{c}& 10\,000 & 3\,495 & 94 & 277 & 310 & - & 29 & - & 184 & - & 167   & 14\,566 \\ \hline
\enddata
\tablenotetext{a}{Later confirm by cross--match.}
\tablenotetext{b}{Total values per class might not represent the direct addition of column values, this is due to some object are present in different catalogs.}
\tablenotetext{c}{Represent the actual total number of sources per class used during training and testing task.}
\end{deluxetable}

\begin{figure*}[ht!]
\gridline{\fig{lc_examples_P.pdf}{0.5\textwidth}{(a)}
          \fig{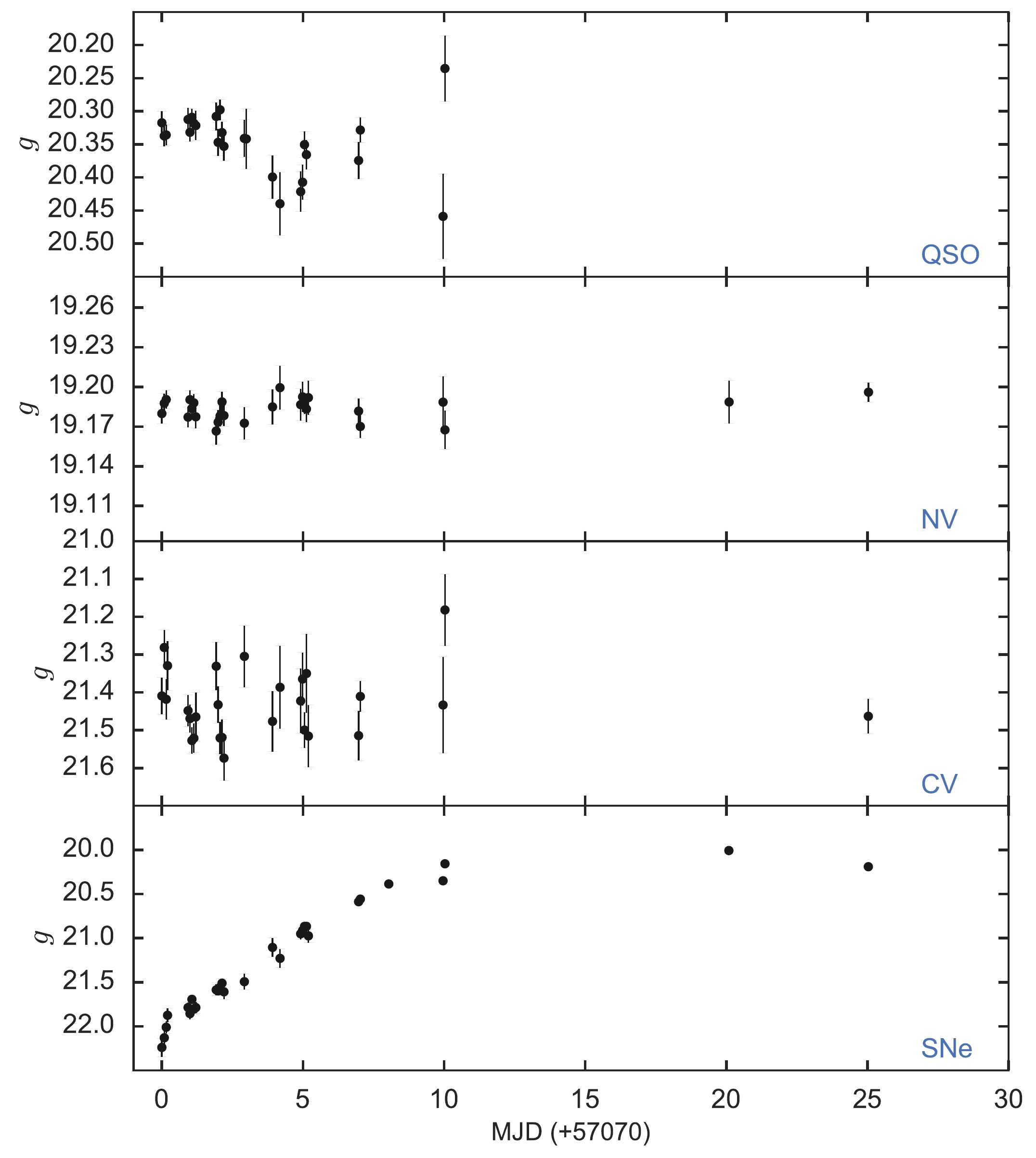}{0.5\textwidth}{(b)}
          }
\caption{(a) Examples of periodic light curves for objects in training sample, light curves are phase folded using period calculation from GLS. (b) Examples of non-periodic light curves for objects in training sample.
\label{figure:LCS}}
\end{figure*}

The total amount of items per class shown in Table \ref{tab:ts_nums} is too small for some classes. For instance, with 1 and 5 items for DSCT and ROTVAR, respectively. In these cases the classification model does not create a good representation of the classes, leading to unreliable results when cross--validation techniques are used. 

In order to compensate the less populated classes we follow a data augmentation approach \citep[see e.g.][]{Dieleman15}. We use known objects in the periodic classes to create synthetic light curves from them by applying basic transformations (i.e. scaling, noise addition and phase shifting). To create these synthetic light curves we follow these steps: we model the observed folded light curves using Gaussian Process \cite[GP,][]{GP}, which is a non--parametric method to model the data. We use the GP regression implemented in \textit{scikit-learn} \footnote{\url{http://scikit-learn.org/stable/modules/gaussian_process.html}} with a periodic \textit{ExpSineSquare}\footnote{Defined as $k(x,x') = \exp\big(- \frac{2 \sin^2(\pi |x-x'|/p)}{l^2}\big)$, where $p$ is the period of the function and $l$ is the length--scale of the function.} kernel, then we sample from this model following the HiTS observing strategy. At this step, we also apply a transformation in the time coordinate to simulate different periods consistent with the distribution of periods in each class; we remove data points in the dim end of the magnitude distribution to emulate the typical fraction of missed data in the light curves; mean values were scaled to the HiTS empirical magnitude distribution, adding heteroscedastic noise using the empirical distributions of errors given a magnitude bin; and finally features were calculated as was done for real light curves.

Following this method, we were able to increase the population for the RRLYR, EB, DSCT and ROTVAR classes (see Table \ref{tab:ts_nums}). For the transient class it is more difficult to perform data augmentation because applying time shifts requires a more difficult interpolation and may require extrapolation. For the non--periodic CV and MISC classes of variable sources it is also difficult to perform data augmentation. CVs exhibit quiescence and outburst behavior in a non--periodic fashion according to their accretion rate with time scales of weeks to years plus timescales of hours due to orbital variations. The MISC class are a heterogeneous family of variable stars and thus difficult to characterize, therefore we eliminated this class from the training set. For periodic ZZ variables, which are fast pulsating stars with periods from seconds to dozens of minutes, even the fast cadence of HiTS is not fast enough for their proper characterization. Therefore, the ZZ class was also removed from the training set.

The final training set contains the classes and numbers listed in Table~\ref{tab:ts_nums} last row. The catalog with sources and classes used for the classification task is available in the project archive of HiTS\textsuperscript{\ref{dr1}}, and the column description of the catalog is presented in the Appendix Section ~\ref{apx: cat_ex} (see Table ~\ref{table:ts_columns}). Examples of periodic and non periodic light curves are shown in Figure \ref{figure:LCS}.

\subsection{Model Training and Testing}

We follow a hierarchical classification scheme using a Random Forest classifier (RF; \citealt{Breiman01}). Briefly, RF consist in a collection of single decision tree classifiers that partition the feature space in a hierarchical fashion, where each tree is trained with a random selection of the objects and the features, and the final classification is the average outcome of each individual tree. RF has been extensively and successfully used in astronomy to classify sources (\citealt{Forster16}, \citealt{Pichara13}, \citealt{Kim14}, \citealt{Yuan17}). For this work we use the  \textit{scikit-learn}\footnote{\url{http://scikit-learn.org/stable/modules/generated/sklearn.ensemble.RandomForestClassifier.html}} implementation of RF. We set the number of estimators (the number of trees) per classifier and the maximum depth of the trees (the maximum path length at which nodes are no longer expanded), minimizing the out-of-bag error  at the training phase (which is the predicted error on a bootstrap aggregation). The remaining hyper--parameters of the classifier (such as the split criteria, the number of features to learn in each tree and the minimum number of samples per leaf) were set as the best F1--score \footnote{The F1--score is the weighted average of the precision and recall, i.e., $F1 = 2 \frac{P*R}{P+R}$, where $P = \frac{T_p}{T_p + F_p}$ is the precision, $\frac{T_p}{T_p + F_n}$ is the recall, and $T_p$ and $F_p$ ($T_n$ and $F_n$) are the numbers of true and false positives (negatives).} value from the cross--validation process. Class imbalance was taken into account weighting each class in the training set by initializing the class weight hyper--parameter as \textit{``balanced\_subsample"}. We fed all the features listed in Table \ref{tab:feat_list} to the RF model. Each classifier in the hierarchical scheme was tested using an unseen set -- the test set -- which represented one third (${\sim} 4\,500$) of the full labeled set, while maintaining imbalance of classes. This set was composed of real sources, with the exception of ROTVAR and DSCT classes due to the small number of real objects.

The hierarchical classification allows us better understanding the possible contaminants of each class.
For the hierarchical scheme we divide the classification into two binary layers and one final multi--class layer. The first layer consists of a binary variable/non--variable classification. The NV class was set as a non--variable class during this step. QSOs were removed from this layer. The reason for this is that short term variability (with time scales of hours) is not a well constrained property of QSOs. Other than Blazars, which are well known fast variables, small samples of a few dozen `normal' QSOs show short term variability in about ${\sim} 10-30\%$ of objects but with amplitudes of ${\sim} 3-10\%$ \citep{Stalin04,Gupta05}. The poorly known properties of this variability could make classification at this stage less reliable.

\begin{figure}[ht!]
\includegraphics[width=.47\textwidth]{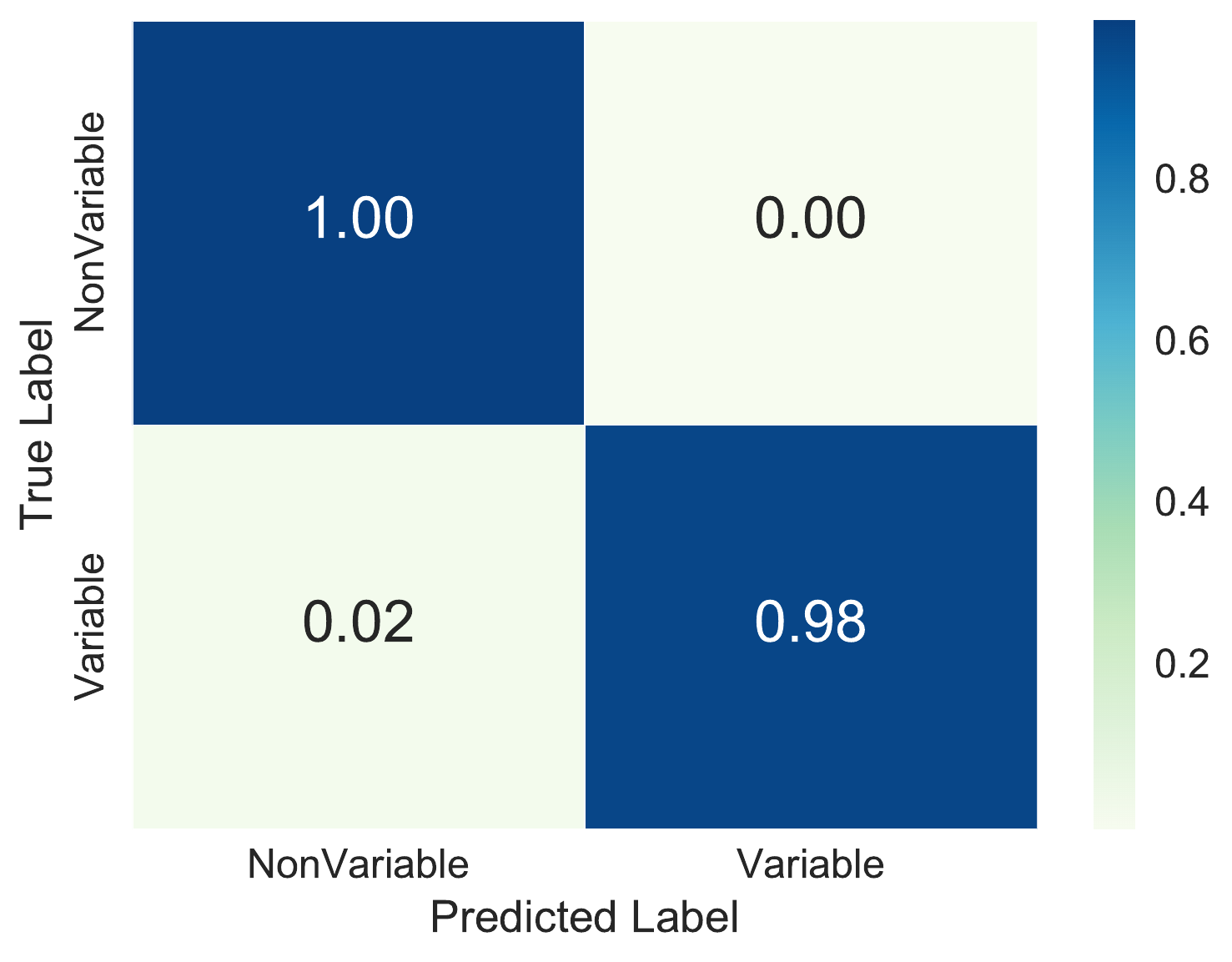}
\caption{Confusion matrix from testing results for variable/non-variable classification. True Label represent the ground truth and Predicted Label the outcome of the RF classifier.
\label{figure:var_nonvar_matrix}}
\end{figure}

A confusion matrix presenting our results is shown in Figure~\ref{figure:var_nonvar_matrix}, where it is easy to see that variables are in general well classified and false positives are statistically zero, i.e. we achieved a high purity. The F1--score on the test set is $99 \pm 0.1\%$ (see Table \ref{table: score_summary} for peer class score values).

\begin{figure}[ht!]

\includegraphics[width=.47\textwidth]{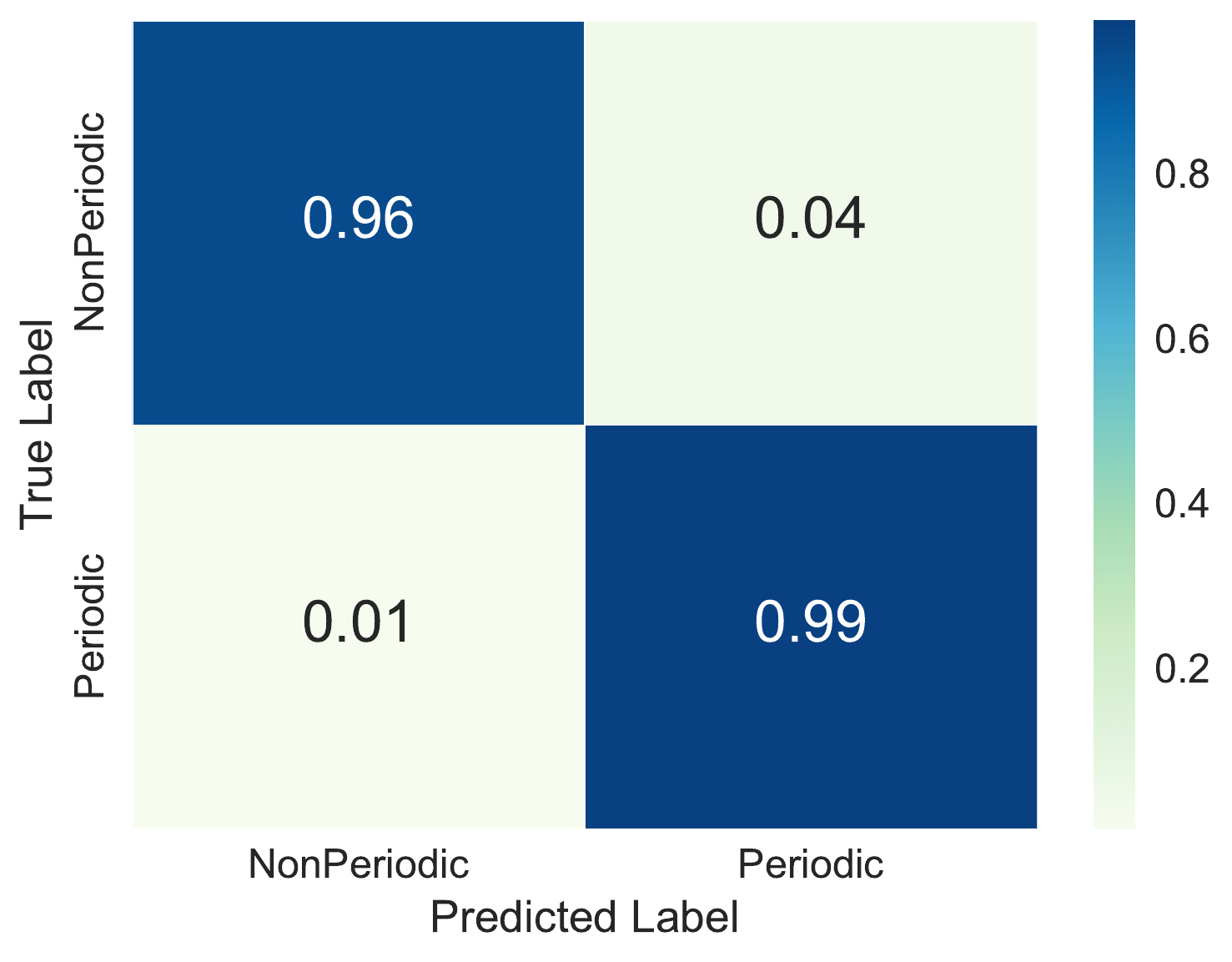}
\caption{Confusion matrix from testing results for periodic/non-periodic classification. Similar to Figure \ref{figure:var_nonvar_matrix}.
\label{figure:per_nonper_matrix}}
\end{figure}

Next, for  variable candidates we separate between periodic and non--periodic classes. Here periodic classes are RRLYR, EB, LPV, ROTVAR and DSCT, while non--periodic classes are CV, SNe and QSOs. Only QSO that were classified as variable with the variable/non--variable classifier (described above) are included in this layer as non--periodic sources, this is 177 out of 3495. We remove all true periodic variables with bad period estimations (i.e. $period\_fit > 0.5$ ) from the training set. The main reason to do this is the presence of periods longer than weeks in the training set, which are not possible to recover with the HiTS observational time span. This only reduces the number of periodics by 68 sources. The F1--score after doing this is $ 98 \pm 0.1 \%$, and the corresponding confusion matrix is shown in Figure~\ref{figure:per_nonper_matrix}. The false positive rate is around $4 \%$, and the classifier misses only $1 \%$ of the periodic sources.

For the final layer, a multi--class classification is applied for each periodic and non-periodic subsets. Within the non--periodic set the classifier was trained with 3 classes (QSO, CV and SNe) and for the periodic set with 4 classes (RRLYR, EB, ROTVAR and DSCT). 

\begin{figure}[ht!]
\includegraphics[width=.47\textwidth]{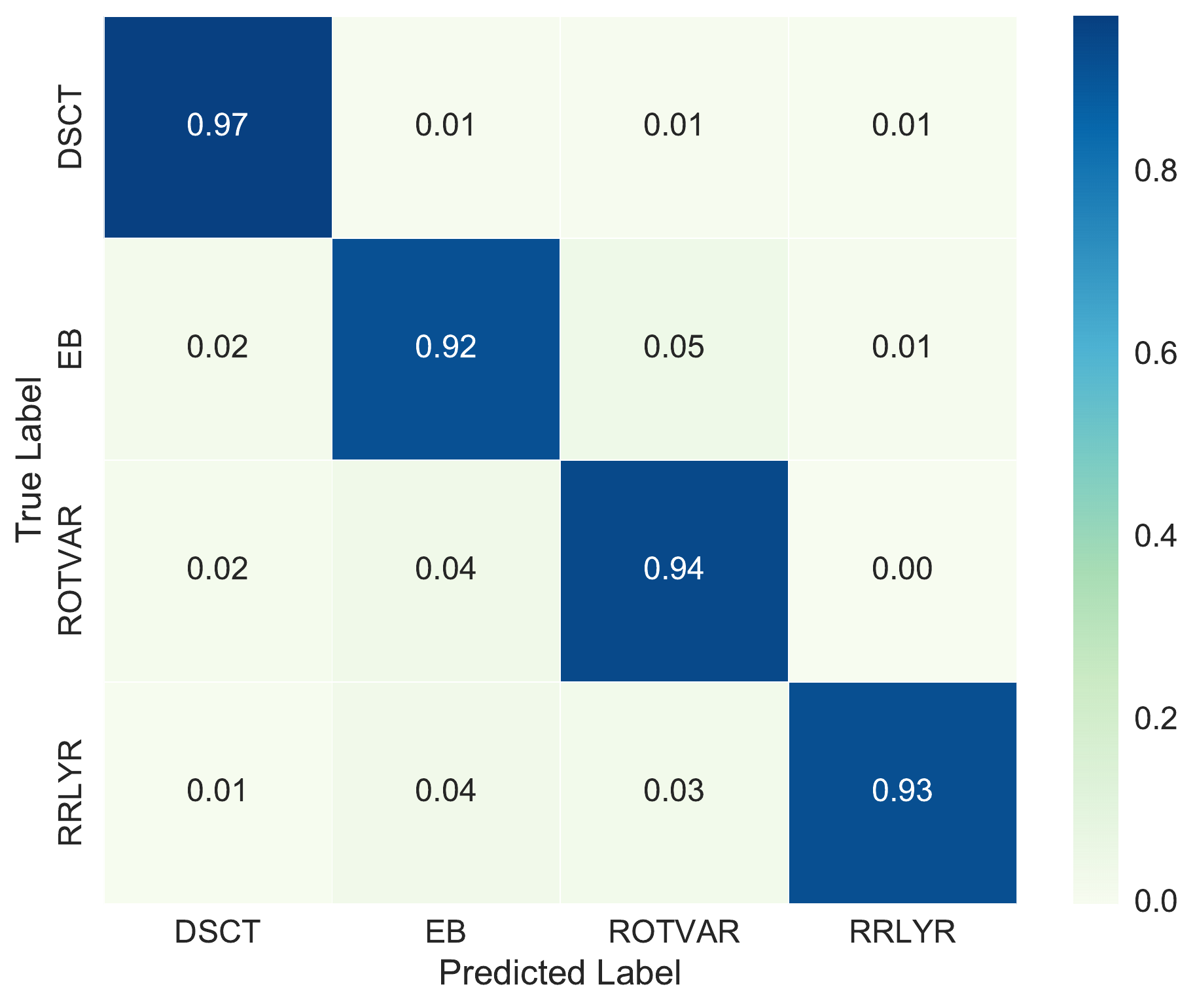}
\caption{Confusion matrix from testing results for periodic sub--classes classification. Similar to Figure \ref{figure:var_nonvar_matrix}.
\label{figure:per_subcls_matrix}}
\end{figure}

For the periodic set classifier, the injection of synthetic light curves significantly improves the performance of the classifier for the ROTVAR and DSCT classes compared with a classifier trained only with real data: from $14\%$ to $94 \%$ and from zero to $97\%$ for ROTVAR and DSCT classes, respectively. However, if we test only with real sources we recover 4 out of 5 sources in the ROTVAR class ($80 \%$ recall), and 1 out of 1 sources in the DSCT class. EB and RRLYR did not show significant improvement after the injection of synthetic light curves. The confusion matrix shown in Figure~\ref{figure:per_subcls_matrix} presents the results on the test set for this layer, where it is possible to notice that in general misclassification is below $5\%$ in all the classes. The weighted F1--score is $93 \pm 2\%$.

\begin{figure}[ht!]
\includegraphics[width=.47\textwidth]{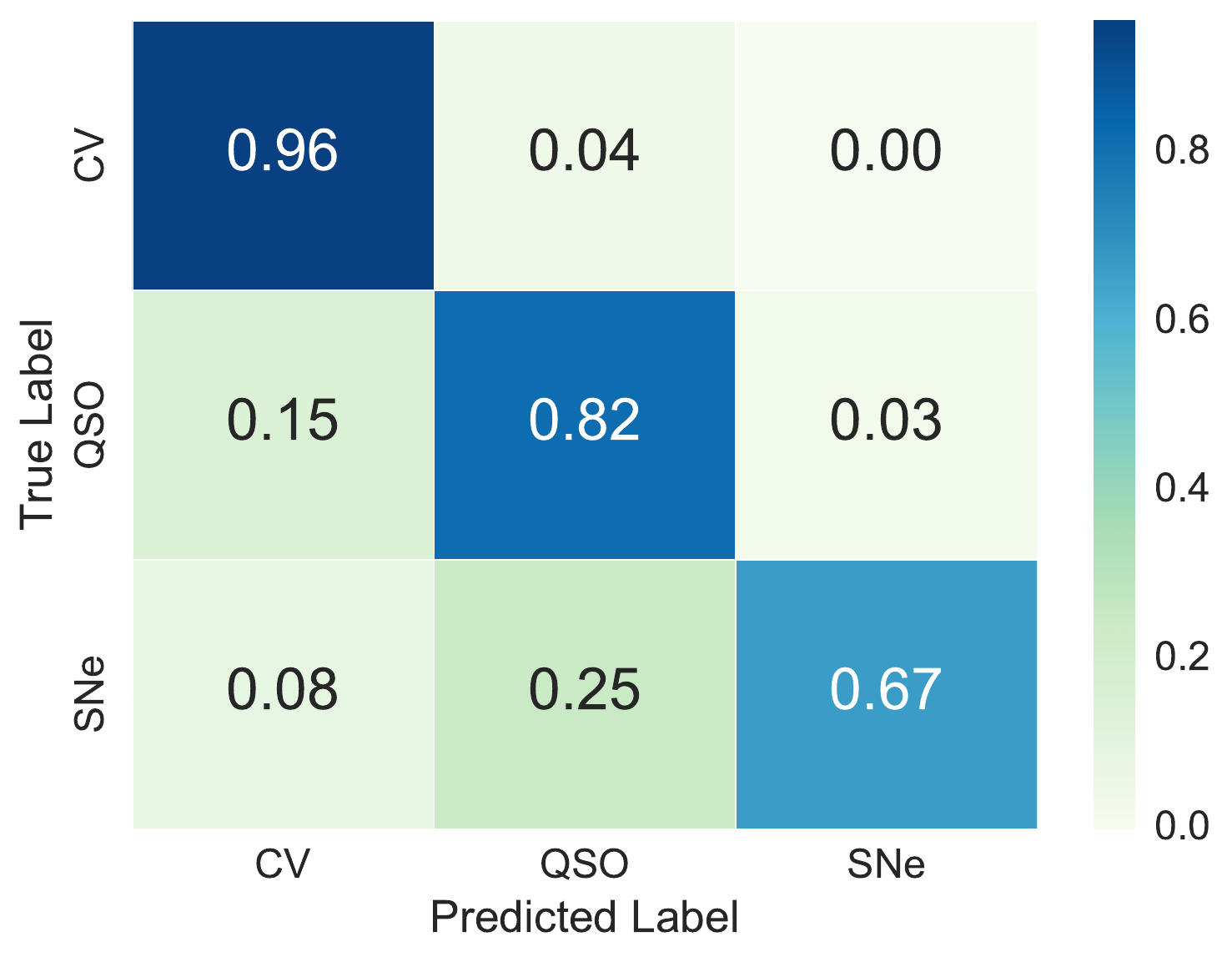}
\caption{Confusion matrix from testing results for non--periodic sub--classes classification. Similar to Figure \ref{figure:var_nonvar_matrix}.
\label{figure:nonper_subcls_matrix}}
\end{figure}

For the non--periodic set the classes are QSO, CV and SNe. The weighted F1--score is $88 \pm 3 \%$. Figure \ref{figure:nonper_subcls_matrix} shows the confusion matrix for this classification model. Here misclassification is higher, given the short time span of the survey. In many cases CVs and QSOs are either observed during their quiescent state or have longer characteristic time scales of variability.

One of the advantages of using RF as a classifier model is that RF naturally provides the importance of each feature as a score. The feature importance reflects which features derived from the light curves separate better each class. The more informative a features is, the higher is its rank. Feature importance ranks for the top ten descriptors are shown in the Appendix Section \ref{apx_sec: feature} (see Table \ref{tab:feat_impor}) for the four classifiers described above. For each classifier, the more informative features are related to the type of classification that is done. In the case of the Variable/non-Variable classification, \textit{Period\_fit} and \textit{Psi\_eta} (variability index for unevenly sampled data) are the most important features, both characterizing the variability in the data. For the Periodic/non--Periodic classification, \textit{Period\_fit} and \textit{CAR\_sigma} are the top two features, the former representing the false alarm probability of the calculated period, and the latter describing the variability dispersion of non--periodic signals. When periodic sources are separated within the four classes present in our training set, the value of the period (\textit{PeriodLS}) is the top ranked discriminator. In the case of the non--Periodic classes, the linear trend (\textit{LinearTrend}) and the characteristic length of the auto--correlation function (\textit{Autocor\_length}), are the most important features.

Table \ref{table: score_summary} summarizes the precision, recall and F1 scores for all the steps in the hierarchical classification described above. We compared our Variable/non--Variable classifier to a classical classification of variability from the standard-deviation--mean plane. For this, we classified as variables all sources with a standard deviation above three times the median standard deviation value for that bin of magnitude. From this we were able to separate variables from non-variables in our data set. We compared the precision, recall and F1 scores of this crude classification (see in Table \ref{table: score_summary} rows labeled as ``basic") against our RF classifier. The RF classifier performed better separating variables. This is because the model uses more information from the set of feature values (see Table \ref{tab:feat_impor} for feature importance derived from the RF classifier) and finds more complex ways to split the data. This demonstrated one of the advantages of using ML algorithms to perform classification of complex datasets.

The final RF classifiers at each step of the hierarchical classification were trained using the full dataset (after cleaning and with the augmented data) with the same initialization conditions as those used during model testing.

\begin{deluxetable}{lccc}[!hbt]
\tablecaption{Summary of precision, recall and F1 scores for every step in the hierarchical classification as measured in the test set. \label{table: score_summary}}
\tablehead{\colhead{Variability Class} & \colhead{Precision} & \colhead{Recall} & \colhead{F1--score}}
\decimals
\startdata
Variable (basic)    &    0.92 & 0.97 & 0.95 \\
NonVariable (basic) &    1 &    0.99 &    0.99 \\
\midrule
Variable (RF)    &    1 & 0.98 & 0.99 \\
NonVariable (RF) &    1 & 1 &    1 \\
\midrule
Periodic    & 0.99 & 0.99 & 0.99 \\
NonPeriodic & 0.97 & 0.96 & 0.96 \\
\midrule
DSCT        & 0.96 & 0.96 & 0.96 \\
EB          & 0.86 & 0.91 & 0.88 \\
ROTVAR      & 0.94 & 0.94 & 0.94 \\
RRLYR       & 0.98 & 0.93 & 0.95 \\
\midrule
CV          & 0.88 & 0.96 & 0.92 \\
QSO         & 0.89 & 0.82 & 0.86 \\
SNe         & 0.80 & 0.67 & 0.73 \\
\enddata
\end{deluxetable}

\section{Results} \label{sec:results}

The first result of this work is a catalog of detected sources for 348 square degrees and a magnitude range between 17 and 24.5 magnitudes in the $u$, $g$, $r$ and $i$ bands. The catalogs for each observation campaign are astrometrically and photometrically corrected, and contain morphology information provided by the extraction process.

A second result is the creation of a training set via cross--matching, active learning, transfer learning and data augmentation. From the latter strategy we are able to fill the entire magnitude range of our survey. 
The final result is a catalog of light curves covering 270 square degrees observed during 2014 and 2015 campaigns. The unlabeled dataset of $2\,536\,100$ sources was classified following the classification scheme previously described. They were first classified as variable/non--variable, then variable candidates were separated between periodic/non--periodic and finally periodic and non--periodic candidates were classified in their respective sub--classes.

\begin{deluxetable}{lrrrrr}[!hbt]
\tabletypesize{\scriptsize}
\tablecaption{Number of predicted sources per variability class, variable and periodic with probability above 90\%, 80\%, 70\%, 60\% and 50\%. Results for all layers in the hierarchical scheme are presented. \label{table: prob_summary}}
\tablehead{\colhead{Class} & \colhead{$>$ 90\%} & \colhead{$>$ 80\%} & \colhead{$>$ 70\%} & \colhead{$>$ 60\%} & \colhead{$>$ 50\%} 
}
\startdata
Variable    &      498   &        822 &      1,361 &      2,130 &      3,485 \\
NonVariable &  2,254,817 &  2,481,443 &  2,520,842 &  2,529,776 &  2,532,615 \\
\midrule
Periodic    &        234 &        397 &        619 &        944 &      1,321 \\
NonPeriodic &        432 &        893 &      1,326 &      1,769 &      2,164 \\
\midrule
DSCT        &          0 &          1 &          5 &         18 &        37 \\
EB          &          6 &         39 &         97 &        176 &       258 \\
ROTVAR      &          8 &         48 &        132 &        296 &       496 \\
RRLYR       &         61 &         90 &        108 &        129 &       142 \\
\midrule
CV          &          0 &          3 &         18 &         79 &       210 \\
QSO         &        160 &        630 &      1,186 &      1,589 &     1,834 \\
SNe         &          0 &          1 &          4 &          6 &        14 \\
\midrule
\enddata
\end{deluxetable}

Classification probabilities from RF are obtained for each layer of the scheme. Table \ref{table: prob_summary} shows the number of predicted sources per variability class. Number counts are shown by probability threshold, i.e., probabilities above 90\%, 80\%, 70\%, 60\% and 50\%. To understand the numbers presented in Table \ref{table: prob_summary} we describe how unlabeled data was classified: first the variable/non--variable model classify variable sources giving a probability where $1$ is variable and $0$ in non--variable, here sources with probability greater than $0.5$ are classified as variables ($3,485$); then this subsample follows the periodic/non--periodic classifier, again with probabilities between $0$ and $1$ where the latter is a periodic signal and sources with probability greater than $0.5$ are classified as periodic, adding the number of periodic and non--periodic sources with probability $> 50\%$ give us the number of variables; then periodic sources go into the periodic classifier which separates between DSCT, EB, RRLYR and ROTVAR, the final class is defined by the one with higher probability from the classifier; same applies to the non--periodic sources that go into the non--periodic classifier in where they were separated between CV, QSO and SNe. Therefore, the numbers presented in the binary periodic/non--periodic classification adds and match the number of variables only in the $> 50\%$ column, but not for higher probability thresholds. For the multiclass classifiers, the addition of numbers is not direct for obvious reasons.

A detailed description of the candidate catalog is presented in the Appendix \ref{apx: cat_ex}, Table \ref{table:var_candidates}. The full catalog contains the outcome probabilities for each layer of the hierarchical classification scheme and it is available along with the source catalog.

In the variable/non--variable layer of the hierarchical model, the model classified a small fraction of the sources as variable. The classifier model found 61 RRLYR, 8 ROTVAR and 6 EB with high confidence, above $90\%$. As comparison, it is estimated that there is one RR--Lyrae per square degree in the MW halo. From there, the expected number of RR--Lyrae in HiTS is roughly 300. Adding the training and candidate RR-Lyrae, this is in agreement with the estimations. On the other hand, from the outcome of the non--periodic layer, $160$ sources are classified as variable QSO candidates with probabilities above $90\%$. Dedicated works of intra--night QSO variability have estimated the fraction of variable sources as $10-30\%$ \citep{Stalin04,Gupta05} and with amplitudes of $3\%$ and $10\%$, respectively. Surface density estimates give ${\sim}20$ and ${\sim}100$ QSOs per square degree at the HiTS \textit{g} band magnitudes of 20 and 22 \citep{Beck07}, at which our photometry is sensitive to variations of $3\%$ and $10\%$, respectively. Therefore a total number of ${\sim}5\,000$ or ${\sim}20\,000$ rapidly varying QSOs are expected in HiTS. Hence, we find a low ratio of variable QSO (from the training and candidate sets) to the total expected number of QSOs (${\sim}7\%$ and ${\sim}1.5\%$). The smaller percentage of variable QSOs might be due to our slow cadence and non--targeted nature of the data, compared to the dedicated works mention above. We will address these intra--night variable QSO in a forthcoming paper.

\subsection{Classification Biases} \label{sec:biases}

The observational strategy used during the HiTS observing campaigns limits our completeness. The most important factors are the magnitude range and the characteristic time scale covered by the survey. The former is represented by the saturation limit of the images which set a lower limit of ${\sim} 15$ magnitude in the $g$ band, and the upper limit set by survey depth calculated as $24.3$ magnitudes in the $g$ band. For comparison, RR--Lyrae at this limiting magnitude will be at $\large 500$ kpc. This is further away than the MW virial radius ${\sim} 300$ kpc, \cite{Barnes17} and therefore it should be complete. More generally, however, both object detection and variability candidate catalogs are restricted to this magnitude range. The characteristic time scale of variable sources studied by this survey is set by the cadence and the maximum time interval of the observations. HiTS time span covers 5/6 consecutive nights, therefore characteristic times greater than weeks are not available. On the other hand, the cadence of HiTS is 2/1.6 hours, making the study of time scales of minutes or seconds very challenging for periodic sources and impossible for fast transients. 

Another set of biases are introduced by the training set at the classification stage. In the ideal case the training set should represent the entire parameter space of the unlabeled data and the number counts per class should represent the true distribution of classes and contain all possible sub--classes. Clearly this is not the case in astronomical studies, where neither all the variability classes nor the number distributions are known. In our case the training set came from cross-matching, active learning and transfer learning and data augmentation technique. It contains 8 variability classes, which are highly unbalanced and they do not necessarily represent the true distribution of variable sources. Additionally, parameters like magnitude distribution per class do not entirely represent the global distribution of sources in the unlabeled data. We minimize these effects by injecting synthetic light curves.

\section{Summary} \label{sec:summary}

The High Cadence Transient Survey was designed for the main purpose of studying the early phase of supernovae events. Despite this, the sky coverage, fast cadence and deep observations make HiTS a survey unique in his type, allowing studies of faint transient and variable sources with characteristic time--scales from hours to days.
In this work, we have compiled a catalog of detections for HiTS, describing the photometric and astrometric calibrations. We estimated limiting and completeness magnitudes of our catalogs, concluding that survey depth is about 24.4, 24.3, 24.1 and 23.8 mag in $u, g, r$ and $i$ bands, respectively. We compared the photometric results with the PanSTARRS catalogs and estimated magnitude deviations of the order of 0.02, 0.04 and 0.07 for g, r and i bands, respectively. Catalogs are available in the project archive of HiTS\textsuperscript{\ref{dr1}}.

We followed a Machine Learning procedure to automatically classify sources by variability. We calculated a set of features designed for variability studies for all the light curves that have more than 15 data points. The total number of light--curves compiled is $2\,536\,100$ from the 2014 and 2015 observation campaigns. We address the difficulties of training set creation following different approaches: the standard procedures of cross--matching with public catalogs (SDSS-DR9, GCVS, CSS and VSX), visual inspections of light curves, or a more Data Science approach such as Active Learning and Transfer Learning. Finally, we took advantage of the known examples for some classes to perform data augmentation.

For the classification model we use RF classifiers following a hierarchical scheme: first, unlabeled data were classified as variable or non--variable; then variable candidates were separated between periodic and non--periodic; then periodic candidates were classified as RRLYR, EB, DSCT and ROTVAR; and non--periodic candidates as QSO, CV and SNe. For each layer of the classification the RF gives classification probabilities that are reported in the variable candidate catalog (the description of the catalog is in Table~\ref{table:var_candidates}).

In this work, we describe different strategies to create the training set that can be followed when supervised classification is used. We present and discuss the problematics that are presented when the training set is built and how this impacts the performance of the classifier. It is important that the characteristic times scales of variables sources that we are searching can be represented with the time span of the data, and the distribution in magnitudes of the training set represents the distribution of the unlabeled data. Otherwise the classifier will perform well only within the range of parameters of the training set. We show how an unbalanced training set leads to a classification that favors more populated classes in the training set and we address this issue injecting synthetic augmented data to classes with periodic signals.

In this work we face the full process of analysis of astronomical data, giving a complete description of the challenges that need to be tackled. We take advantage of the unique characteristics of th HiTS survey such as survey depth, field of view, observation cadence and data volume, which presents an excellent laboratory for the next generation of large surveys like LSST. HiTS observations reach a similar limiting magnitude than LSST, therefore it provides an important data set to test LSST future software, and to start building libraries of light curves for different variability classes which will be extremely important for the process of supervised automatic classifications needed to analyze LSST data.

Finally, the catalogs released by this work represent the first Data Release (DR1) of the HiTS survey. For the next DR we plan to implement PSF photometry, as well as forced photometry for sources with no detections in a given epoch image. Additionally, we will improve the data access using a database framework, for which we are testing the use of a dedicated time series database like influxDB or others.

\acknowledgments
J.M. acknowledges the support from CONICYT Chile through (CONICYT-PCHA/Doctorado-Nacional/2014-21140892).
J.M., F.F., G.C.V. and G.M. acknowledges the support from Ministry of Economy, Development, and Tourism's Millennium Science Initiative through grant IC120009, awarded to The Millennium Institute of Astrophysics (MAS).
F.F. acknowledges support from Conicyt through the Fondecyt Initiation into Research project No. 11130228.
J.M., F.F., J.S.M., G.C.V., and S.G. acknowledge support from Basal Project PFB-03, Centro de Modelamiento Matemáico (CMM), Universidad de Chile.
P.L. acknowledges support by Fondecyt through Project \#1161184.
G.C.V. gratefully acknowledges financial support from CONICYT-Chile through its FONDECYT postdoctoral grant number 3160747; CONICYT-Chile and NSF through the Programme of International Cooperation project DPI201400090.
P.H. acknowledges support from FONDECYT through grant 1170305.
L.G. was supported in part by the US National Science Foundation under Grant AST-1311862.
G.M. acknowledges support from Conicyt through CONICYT-PCHA/Mag\'isterNacional/2016-22162353.
Support for T.d.J. has been provided by US NSF grant AST-1211916, the TABASGO Foundation, Gary and Cynthia Bengier.
R.~R.~M.~acknowledges partial support from BASAL Project PFB-$06$ as well as FONDECYT project N$^{\circ}1170364$.
Powered@NLHPC: this research was supported by the High Performance Computing infrastructure of the National Laboratory for High Performance Computing (NLHPC). PIA ECM-02. CONICYT.
This project used data obtained with the Dark Energy Camera (DECam), which was constructed by the Dark Energy Survey (DES) collaborating institutions: Argonne National Lab, University of California Santa Cruz, University of Cambridge, Centro de Investigaciones Energeticas, Medioambientales y Tecnologicas-Madrid, University of Chicago, University College London, DES-Brazil consortium, University of Edinburgh, ETH-Zurich, University of Illinois at Urbana-Champaign, Institut de Ciencies de l'Espai, Institut de Fisica d'Altes Energies, Lawrence Berkeley National Lab, Ludwig-Maximilians Universitat, University of Michigan, National Optical Astronomy Observatory, University of Nottingham, Ohio State University, University of Pennsylvania, University of Portsmouth, SLAC National Lab, Stanford University, University of Sussex, and Texas A\&M University. Funding for DES, including DECam, has been provided by the U.S. Department of Energy, National Science Foundation, Ministry of Education and Science (Spain), Science and Technology Facilities Council (UK), Higher Education Funding Council (England), National Center for Supercomputing Applications, Kavli Institute for Cosmological Physics, Financiadora de Estudos e Projetos, Fundação Carlos Chagas Filho de Amparo a Pesquisa, Conselho Nacional de Desenvolvimento Científico e Tecnológico and the Ministério da Ciência e Tecnologia (Brazil), the German Research Foundation-sponsored cluster of excellence "Origin and Structure of the Universe" and the DES collaborating institutions.

\vspace{5mm}
\facility{CTIO:1.5m(DECam)}

\software{Crblaster \citep{Mighell10}, 
          SExtractor \citep{Bertin10},
          SCAMP \citep{SCAMP},
          scikit-learn \citep{scikit-learn}}



\newpage

\appendix
\section{Catalog Description} \label{apx: cat_ex}
\vspace{-0.8cm}
\begin{deluxetable}{llll}[!ht]
\tabletypesize{\scriptsize}
\tablecaption{List of columns names, units and description for object catalog. Tables for 2013 and 2014 data only have one observational band, u and g, respectively. \label{table:catalog_cols}}
\tablehead{\colhead{Column Name} & \colhead{Units} & \colhead{Default Value} & \colhead{Description}} 
\startdata
ID	&  dimensionless  & NA &  Survey ID of the source ("HiTS"+"hhmmss"+"ddmmss") \\
internalID	&  dimensionless  & NA &  Internal ID of the source ("Filed"\_"CCD"\_"Xpix"\_"Ypix") \\
X	&  pixels  & -999 &  Pixel X position in reference image \\
Y	&  pixels  & -999 &  Pixel Y position in reference image \\
raMedian	&  degrees  & -999 &  Median Right Ascension position  \\
decMedian	&  degrees  & -999 &  Median Declination position \\
raMedianStd	&  degrees  & -999 &  Standard deviation of Right Ascension across detections \\
decMedianStd	&  degrees  & -999 &  Standard deviation of Declination across detections \\
ugri\tablenotemark{*}N  &  dimensionless  & 0 &  Number of single detection in a given band band \\
ugri\tablenotemark{*}ClassStar  &  dimensionless  & 0 &  Galaxy/Star classification from SExtractor\tablenotemark{a} \\
ugri\tablenotemark{*}Ellipticity  &  dimensionless  & -999 &  Derived ellipticity from SExtractor's Kron aperture  \\
ugri\tablenotemark{*}FWHM  &  arcsec  & -999 &  SExtractor's FWHM for the source \\
ugri\tablenotemark{*}Flags  &  dimensionless  & 0 &  SExtractor's Flag\tablenotemark{a} for the source \\
ugri\tablenotemark{*}FluxRadius  &  arcsec  & -999 &  Radius at 50\% of the total flux \\
ugri\tablenotemark{*}KronRadius  &  arcsec  & -999 &  Kron radius for a 2D aperture \tablenotemark{a} \\
ugri\tablenotemark{*}MedianAp1Flux  &  $\mu$Jy  & -999 &  Median integrated flux within circular aperture with 1 FWHM radius \\
ugri\tablenotemark{*}MedianAp1FluxErr  &  $\mu$Jy  & -999 &  Median error of the integrated flux within circular aperture with 1 FWHM radius \\
ugri\tablenotemark{*}MedianAp1FluxStd  &  $\mu$Jy  & -999 &  Standard deviation of integrated flux within circular aperture with 1 FWHM radius \\
ugri\tablenotemark{*}MedianAp1Mag  &  AB magnitudes  & -999 &  Median magnitude within circular aperture with 1 FWHM radius \\
ugri\tablenotemark{*}MedianAp1MagErr  &  AB magnitudes  & -999 &  Median error of the magnitude within circular aperture with 1 FWHM radius \\
ugri\tablenotemark{*}MedianAp1MagStd  &  AB magnitudes  & -999 &  Standard deviation of the magnitude within circular aperture with 1 FWHM radius \\
ugri\tablenotemark{*}MedianAp2Flux  &  $\mu$Jy  & -999 &  Median integrated flux within circular aperture with 2 FWHM radius \\
ugri\tablenotemark{*}MedianAp2FluxErr  &  $\mu$Jy  & -999 &  Median error of the integrated flux within circular aperture with 2 FWHM radius \\
ugri\tablenotemark{*}MedianAp2FluxStd  &  $\mu$Jy  & -999 &  Standard deviation of integrated flux within circular aperture with 2 FWHM radius \\
ugri\tablenotemark{*}MedianAp2Mag  &  AB magnitudes  & -999 &  Median magnitude within circular aperture with 2 FWHM radius \\
ugri\tablenotemark{*}MedianAp2MagErr  &  AB magnitudes  & -999 &  Median error of the magnitude within circular aperture with 2 FWHM radius \\
ugri\tablenotemark{*}MedianAp2MagStd  &  AB magnitudes  & -999 &  Standard deviation of the magnitude within circular aperture with 2 FWHM radius \\
ugri\tablenotemark{*}MedianAp3Flux  &  $\mu$Jy  & -999 &  Median integrated flux within circular aperture with 3 FWHM radius \\
ugri\tablenotemark{*}MedianAp3FluxErr  &  $\mu$Jy  & -999 &  Median error of the integrated flux within circular aperture with 3 FWHM radius \\
ugri\tablenotemark{*}MedianAp3FluxStd  &  $\mu$Jy  & -999 &  Standard deviation of integrated flux within circular aperture with 3 FWHM radius \\
ugri\tablenotemark{*}MedianAp3Mag  &  AB magnitudes  & -999 &  Mediand magnitude within circular aperture with 3 FWHM radius \\
ugri\tablenotemark{*}MedianAp3MagErr  &  AB magnitudes  & -999 &  Median error of the magnitude within circular aperture with 3 FWHM radius \\
ugri\tablenotemark{*}MedianAp3MagStd  &  AB magnitudes  & -999 &  Standard deviation of the magnitude within circular aperture with 3 FWHM radius \\
ugri\tablenotemark{*}MedianAp4Flux  &  $\mu$Jy  & -999 &  Median integrated flux within circular aperture with 4 FWHM radius \\
ugri\tablenotemark{*}MedianAp4FluxErr  &  $\mu$Jy  & -999 &  Median error of the integrated flux within circular aperture with 4 FWHM radius \\
ugri\tablenotemark{*}MedianAp4FluxStd  &  $\mu$Jy  & -999 &  Standard deviation of integrated flux within circular aperture with 4 FWHM radius \\
ugri\tablenotemark{*}MedianAp4Mag  &  AB magnitudes  & -999 &  Median magnitude within circular aperture with 4 FWHM radius \\
ugri\tablenotemark{*}MedianAp4MagErr  &  AB magnitudes  & -999 &  Median error of the magnitude within circular aperture with 4 FWHM radius \\
ugri\tablenotemark{*}MedianAp4MagStd  &  AB magnitudes  & -999 &  Standard deviation of the magnitude within circular aperture with 4 FWHM radius \\
ugri\tablenotemark{*}MedianAp5Flux  &  $\mu$Jy  & -999 &  Median integrated flux within circular aperture with 5 FWHM radius \\
ugri\tablenotemark{*}MedianAp5FluxErr  &  $\mu$Jy  & -999 &  Median error of the integrated flux within circular aperture with 5 FWHM radius \\
ugri\tablenotemark{*}MedianAp5FluxStd  &  $\mu$Jy  & -999 &  Standard deviation of integrated flux within circular aperture with 5 FWHM radius \\
ugri\tablenotemark{*}MedianAp5Mag  &  AB magnitudes  & -999 &  Median magnitude within circular aperture with 5 FWHM radius \\
ugri\tablenotemark{*}MedianAp5MagErr  &  AB magnitudes  & -999 &  Median error of the magnitude within circular aperture with 5 FWHM radius \\
ugri\tablenotemark{*}MedianAp5MagStd  &  AB magnitudes  & -999 &  Standard deviation of the magnitude within circular aperture with 5 FWHM radius \\
ugri\tablenotemark{*}MedianKronFlux  &  $\mu$Jy  & -999 &  Median integrated flux within Kron aperture \\
ugri\tablenotemark{*}MedianKronFluxErr  &  $\mu$Jy  & -999 &  Median error of the integrated flux within Kron aperture \\
ugri\tablenotemark{*}MedianKronFluxStd  &  $\mu$Jy  & -999 &  Standard deviation of integrated flux within Kron aperture \\
ugri\tablenotemark{*}MedianKronMag  &  AB magnitudes  & -999 &  Median magnitude within Kron aperture \\
ugri\tablenotemark{*}MedianKronMagErr  &  AB magnitudes  & -999 &  Median error of the magnitude within Kron aperture \\
ugri\tablenotemark{*}MedianKronMagStd  &  AB magnitudes  & -999 &  Standard deviation of the magnitude within Kron aperture \\
\enddata
\tablenotetext{*}{All columns are compute for $u,g,r$ and $i$ bands if image data is available.}
\tablenotetext{a}{See \cite{Bertin10} for further information.}
\end{deluxetable}


\begin{deluxetable}{llll}[!ht]
\tabletypesize{\normalsize}
\tablecaption{List of columns names, units and description for the training set catalog used during the training and testing phase. \label{table:ts_columns}}
\tablehead{\colhead{Column Name} & \colhead{Units} & \colhead{Default Value} & \colhead{Description}} 
\startdata
ID	&  dimensionless  & NA &  Survey ID of the source ("HiTS"+"hhmmss"+"sign"+"ddmmss") \\
internalID	&  dimensionless  & NA &  Internal ID of the source ("Filed"\_"CCD"\_"Xpix"\_"Ypix") \\
raMedian	&  degrees  & -999 &  Median Right Ascension position  \\
decMedian	&  degrees  & -999 &  Median Declination position \\
spCl		&  dimensionless   & NA   &  Spectral class from SDSS-DR9 cross--match result\tablenotemark{a} \\
spSubCl		&  dimensionless   & NA   &  Spectral subclass from SDSS-DR9 cross--match result\tablenotemark{a} \\
Var\_Type	&  dimensionless   & NA   &  Variability class \\
Var\_subType	&  dimensionless   & NA   &  Variability subclass \\
Augmented\_data &  dimensionless & -999 & Int flag, 0 is real data and 1 is synthetic data \\
\enddata
\tablenotetext{a}{See \cite{SDSSDR9_BOSS} for further information.}
\end{deluxetable}


\begin{deluxetable}{llll}[!hp]
\tabletypesize{\normalsize}
\tablecaption{List of columns names, units and description for variable candidates catalog, the classification result from the hierarchical RF. \label{table:var_candidates}}
\tablehead{\colhead{Column Name} & \colhead{Units} & \colhead{Default Value} & \colhead{Description}} 
\startdata
ID	&  dimensionless  & NA &  Survey ID of the source ("HiTS"+"hhmmss"+"ddmmss") \\
internalID	&  dimensionless  & NA &  Internal ID of the source ("Filed"\_"CCD"\_"Xpix"\_"Ypix") \\
raMedian	&  degrees  & -999 &  Median Right Ascension position  \\
decMedian	&  degrees  & -999 &  Median Declination position \\
Variable\_Prob  & dimensionless  & NA & Classification probability from the variable/non--variable layer \\
Periodic\_Prob  & dimensionless  & NA & Classification probability from the periodic/non--periodic layer \\
DSCT\_Prob  & dimensionless  & NA & Classification probability from the periodic sub--classes \\
EB\_Prob  & dimensionless  & NA & Classification probability from the periodic sub--classes \\
ROTVAR\_Prob  & dimensionless  & NA & Classification probability from the periodic sub--classes \\
RRLYR\_Prob  & dimensionless  & NA & Classification probability from the periodic sub--classes \\
CV\_Prob  & dimensionless  & NA & Classification probability from the nonperiodic sub--classes \\
QSO\_Prob  & dimensionless  & NA & Classification probability from the nonperiodic sub--classes \\
SNe\_Prob  & dimensionless  & NA & Classification probability from the nonperiodic sub--classes \\
Predicted\_Class & string & NA & Final label for the classification task \\
\enddata
\end{deluxetable}

\clearpage
\section{Feature List and Importance} \label{apx_sec: feature}

\begin{deluxetable}{lll}[!hp]
\tabletypesize{\normalsize}
\tablecaption{List of features used in this work, we use features previously defined in FATS library and we add period estimation from GLS and CKP calculated by \textit{gatspy} and \textit{P4j} python packages, respectively. Color indexes were calculated from g, r and i bands. For further details see \cite{Nun15}. \label{tab:feat_list}}
\tablehead{
\colhead{Feature} & \colhead{Feature} & \colhead{Feature}}
\startdata
Amplitude                         & Freq2\_harmonics\_amplitude\_0    &  MedianBRP                        \\
AndersonDarling                   & Freq2\_harmonics\_amplitude\_1    &  PairSlopeTrend                   \\
Autocor\_length                   & Freq2\_harmonics\_amplitude\_2    &  PercentAmplitude                 \\
Beyond1Std                        & Freq2\_harmonics\_amplitude\_3    &  PercentDifferenceFluxPercentile  \\
CAR\_mean                         & Freq2\_harmonics\_rel\_phase\_0   &  PeriodGLS                        \\
CAR\_sigma                        & Freq2\_harmonics\_rel\_phase\_1   &  PeriodLS                         \\
CAR\_tau                          & Freq2\_harmonics\_rel\_phase\_2   &  PeriodWMCC                       \\
Con                               & Freq2\_harmonics\_rel\_phase\_3   &  Period\_fit                      \\
Eta\_e                            & Freq3\_harmonics\_amplitude\_0    &  Psi\_CS                          \\
FluxPercentileRatioMid20          & Freq3\_harmonics\_amplitude\_1    &  Psi\_eta                         \\
FluxPercentileRatioMid35          & Freq3\_harmonics\_amplitude\_2    &  Q31                              \\
FluxPercentileRatioMid50          & Freq3\_harmonics\_amplitude\_3    &  Rcs                              \\
FluxPercentileRatioMid65          & Freq3\_harmonics\_rel\_phase\_0   &  Skew                             \\
FluxPercentileRatioMid80          & Freq3\_harmonics\_rel\_phase\_1   &  SlottedA\_length                 \\
Freq1\_harmonics\_amplitude\_0    & Freq3\_harmonics\_rel\_phase\_2   &  SmallKurtosis                    \\
Freq1\_harmonics\_amplitude\_1    & Freq3\_harmonics\_rel\_phase\_3   &  Std                              \\
Freq1\_harmonics\_amplitude\_2    & Gskew                             &  StetsonK                         \\
Freq1\_harmonics\_amplitude\_3    & LinearTrend                       &  StetsonK\_AC                     \\
Freq1\_harmonics\_rel\_phase\_0   & MaxSlope                          &  g--i                             \\
Freq1\_harmonics\_rel\_phase\_1   & Mean                              &  g--r                             \\
Freq1\_harmonics\_rel\_phase\_2   & Meanvariance                      &  r--i                             \\
Freq1\_harmonics\_rel\_phase\_3   & MedianAbsDev                      &                                   \\
\enddata
\end{deluxetable}

\clearpage

\begin{splitdeluxetable}{llllBllll}
\tabletypesize{\normalsize}
\tablecaption{Feature importance (top 10) derived from RF classifier at each layer of the hierarchical scheme. First two columns refers to the variable/non--variable classifier, then the next 2 columns to the periodic/non--periodic, columns (5) and (6) refers to periodic classes (DSCT, EB, ROTVAR and RRLYR) and the final two columns to non--periodic classes (CV, QSO and SNe). \label{tab:feat_impor}}
\tablewidth{0pt}
\decimalcolnumbers
\tablehead{
\multicolumn{2}{c}{Variable/non--Variable} & \multicolumn{2}{c}{Periodic/non--Periodic} &
\multicolumn{2}{c}{Periodic classes} & \multicolumn{2}{c}{non--Periodic classes}\\
\cmidrule(r){1-2} \cmidrule(r){3-4} \cmidrule(r){5-6} \cmidrule(r){7-8}
\colhead{Feature} & \colhead{Rank} & \colhead{Feature} & \colhead{Rank} & \colhead{Feature} & \colhead{Rank} & \colhead{Feature} & \colhead{Rank}}
\startdata
Period\_fit & 0.1813 & Period\_fit & 0.1837 & PeriodLS & 0.0912 & LinearTrend & 0.0734 \\
Psi\_eta & 0.1417 & CAR\_sigma & 0.1133 & CAR\_tau & 0.0823 & Autocor\_length & 0.0679 \\
SmallKurtosis & 0.0999 & Freq1\_harmonics\_amplitude\_0 & 0.0869 & CAR\_sigma & 0.0769 & Amplitude & 0.0514 \\
StetsonK\_AC & 0.0671 & Psi\_eta & 0.0860 & Gskew & 0.0757 & Psi\_eta & 0.0489 \\
Mean & 0.0572 & Std & 0.0784 & Meanvariance & 0.0723 & Rcs & 0.0448 \\
PeriodLS & 0.0509 & Q31 & 0.0610 & CAR\_mean & 0.0690 & PeriodLS & 0.0384 \\
Q31 & 0.0496 & Meanvariance & 0.0461 & Skew & 0.0498 & Freq1\_harmonics\_amplitude\_3 & 0.0371 \\
Meanvariance & 0.0492 & CAR\_tau & 0.0322 & PercentDifferenceFluxPercentile & 0.0444 & Meanvariance & 0.0365 \\
MedianAbsDev & 0.0469 & PeriodLS & 0.0251 & PeriodGLS & 0.0400 & Q31 & 0.0358 \\
PercentDifferenceFluxPercentile & 0.0419 & SmallKurtosis & 0.0207 & MaxSlope & 0.0390 & Freq1\_harmonics\_amplitude\_1 & 0.0346 \\
\enddata
\end{splitdeluxetable}

\clearpage
\bibliography{bibliography.bib}



\end{document}